\documentclass[10pt,aps,pra,twocolumn,showpacs,superscriptaddress]{revtex4-1}  % for review and submission
\usepackage{array,longtable}
\newcolumntype{C}{>{$}c<{$}}  % automatic math mode, centered
\usepackage{siunitx}
\usepackage[utf8]{inputenc}
\usepackage{graphicx}
\usepackage{amsmath}
\usepackage{amssymb}
\allowdisplaybreaks[4]
\usepackage{bm}        % for math
\usepackage{txfonts}
\usepackage{mathptmx}  % Times font
\usepackage{siunitx}
\usepackage{comment}
\usepackage{xargs}                      % Use more than one optional parameter in a new commands
\usepackage[pdftex,dvipsnames]{xcolor}  % Coloured text etc.
\usepackage[colorinlistoftodos,prependcaption,textsize=meidum]{todonotes}

\newcommand{\ket}[1]{\left| #1 \right\rangle}
\usepackage[%
  colorlinks=true,
  urlcolor=blue,
  linkcolor=blue,
  citecolor=blue
]{hyperref}

\begin{document}
\title{Multipartite Entanglement for Multi-node Quantum Networks}
\author{E. M. Ainley}
\author{A. Agrawal}
\author{D. Main}
\author{P. Drmota}
\author{D. P. Nadlinger}
\author{B. C. Nichol}
\author{R. Srinivas}
\author{G. Araneda}
\altaffiliation{Corresponding author}
\email{gabriel.aranedamachuca@physics.ox.ac.uk}
\affiliation{Department of Physics, University of Oxford, Clarendon Laboratory, Parks Road, Oxford OX1 3PU, United Kingdom}

\date{\today}
\begin{abstract}
Scaling the number of entangled nodes in a quantum network is a challenge with significant implications for quantum computing, clock synchronisation, secure communications, and quantum sensing. In a quantum network, photons interact with matter qubits at different nodes, flexibly enabling the creation of remote entanglement between them. Multipartite entanglement among multiple nodes will be crucial for many proposed quantum network applications, including quantum computational tasks and quantum metrology. To date, experimental efforts have primarily focused on generating bipartite entanglement between nodes, which is widely regarded as the fundamental quantum resource for quantum networks. However, relying exclusively on bipartite entanglement to form more complex multipartite entanglement introduces several challenges. These include the need for ancillary qubits, extensive local entangling operations which increases the preparation latency, and increasingly stringent requirements on coherence times as the number of nodes grows. Here, we analyse various schemes that achieve multipartite entanglement between nodes in a single step, bypassing the need for multiple rounds of bipartite entanglement. We demonstrate that different schemes can produce distinct multipartite entangled states, with varying fidelity and generation rates. Additionally, we discuss the applicability of these schemes across different experimental platforms, highlighting their primary advantages and disadvantages.
\end{abstract}
\maketitle

\tableofcontents

\section{Introduction}
The need for quantum networks is ubiquitous in the field of quantum technologies and fundamental quantum physics. Multi-node quantum networks have been proposed to compare and synchronise atomic clocks, or to create sensing networks that could enrich our understanding of the universe \cite{derevianko2022quantum}. In fundamental quantum physics, they will permit testing concepts such as local realism to a new extent \cite{greenberger1990bell, dur2001multipartite}. In quantum communications, multi-node networks will enable essentially secure multi-party key exchange \cite{holz2020genuine}. In quantum computing, the ability to engineer complex and high-fidelity entangled states between separated processors will allow for the construction of modular quantum computers \cite{grover1997quantum,cirac1999distributed}. Increasing the number of processing nodes and qubits while maintaining a high degree of connectivity between all qubits is crucial for practical and useful quantum computation \cite{yuan2023does}.

So far, bipartite entanglement between nodes mediated by photons has generally been considered as the initial quantum network resource, across all experimental platforms \cite{eisert2000optimal,oi2006scalable,jiang2007distributed,gottesman1999quantum,avis2023analysis}. Nevertheless, most applications require more complex multipartite entangled states, such as GHZ, W, graph, and cluster states \cite{eisert2005multi, komar2014quantum,hillery1999quantum, greenberger1990bell,chen2011spin,graham2015superdense,gottesman1999quantum,liu2011efficient,joo2002quantum}. Building multipartite entanglement from bipartite entanglement relies on using multiple entangled pairs, local mid-circuit measurements, and feed-forward control of the qubits concerned \cite{ferrari2023modular,avis2023analysis}; these operations are costly in time and resources and increase network latency. Importantly, it has been shown that no graph state can be created over a quantum network without classical communication \cite{makuta2023no,wang2024quantum}, which makes this latency unavoidable. The network latency not only increases the duration of quantum computations, but also, in realistic physical implementations, means that the time required to generate some particular multipartite entangled state can take longer than the coherence time of the individual qubits, making the computation impossible without the need of additional memory qubits. Although long-lived qubits have been demonstrated across different platforms, the qubits used for computation are not necessarily the most adequate to interface with the photonic network, adding error overheads when entanglement needs to be swapped between network and processing qubits (see e.g., \cite{drmota2023robust}).

Hence, it is natural to look for quantum network resources beyond bipartite entanglement. In this article, we examine different schemes experimentally considered so far to create bipartite entanglement mediated by photons in different experimental platforms, and extend them to multiple parties. We identify the types of multipartite entanglement produced, and discuss their main limitations and strengths. All the schemes that we consider create multipartite entanglement \textit{in a single shot}, i.e., they are processes that require at most one round of classical communication and feed-forward, and do not rely on bipartite remote entanglement and local entangling operations to create more complex entangled states. %Some of this scheme have been previously considered \cite{valivarthi2014efficient, wang2009schemes}.

\begin{figure*}
\centerline{\includegraphics[width=0.85\textwidth]{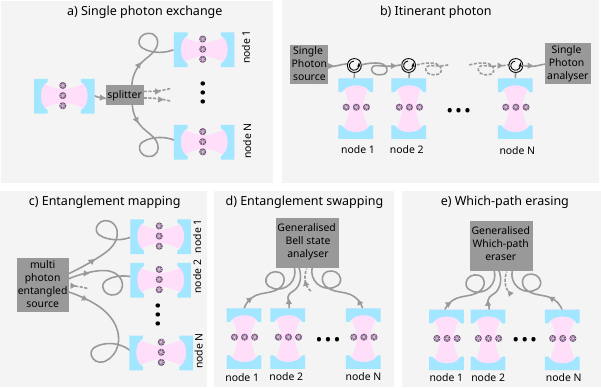}}
\caption{Entangling schemes enabling generation of multipartite entanglement between multiple nodes. \textbf{a)} Photon exchange. The state of one of the atoms is mapped to the state of a photon, which is then routed to different nodes, where its state is mapped to the state of one atom. \textbf{b)} A passing (\textit{itinerant}) photon emitted by a single-photon source (SPS) interacts with different nodes, implementing an entangling gate with one atom in each of the nodes; then the photon is analysed in a single-photon analyser (SPA). \textbf{c)} A multi-photon entangled source (MPES) is used to create several entangled photons. The states of the photons are then mapped to the states of the atoms in different nodes. \textbf{d)} Entangled atom-photon pairs are generated. The photons enter a generalised Bell state analyser (GBSA), and upon detection of photons in coincidence, the states of the atoms are projected into multipartite entangled states. \textbf{e)} One or more photons are emitted collectively and synchronously from different nodes. The photons enter a generalised which-path-eraser (GWPE). The detection of an indistinguishable photon projects the atoms concerned into a multipartite entangled state.}
\label{fig:schemes}
\end{figure*}

The various schemes considered in this article are summarised in Fig.\,\ref{fig:schemes}. Fig.\,\ref{fig:schemes}a shows the ``photon exchange'' scheme~\cite{cirac1997quantum}, Fig.\,\ref{fig:schemes}b shows the ``itinerant photon'' scheme~\cite{duan2004scalable}, Fig.\,\ref{fig:schemes}c shows the ``photon-to-atom mapping'' scheme~\cite{sangouard2013heralded}, Fig.\,\ref{fig:schemes}d shows the ``entanglement swapping'' scheme~\cite{feng2003entangling,duan2003efficient,simon2003robust}, and Fig.\ \ref{fig:schemes}e shows the ``which-path erasing'' scheme~\cite{cabrillo1999creation}.
In each case, each node contains one or several quantum emitters. These quantum emitters can be, for example, single trapped ions \cite{moehring2007entanglement,krutyanskiy2023entanglement}, single neutral atoms \cite{ritter2012elementary,hofmann2012heralded}, clouds of neutral atoms \cite{chou2005measurement,bao2012quantum}, superconducting qubits \cite{magnard2020microwave,campagne2018deterministic}, crystal defects \cite{humphreys2018deterministic, knaut2024entanglement}, quantum dots \cite{delteil2016generation,stockill2017phase}, rare-earth ions in a crystal \cite{ruskuc2024scalable, maring2017photonic}, silicon T-centres \cite{higginbottom2022optical}, among others. For simplicity, we refer to all these as ``atoms" throughout this article. These atoms can emit single photons, which can then be captured either by using a resonator \cite{reiserer2015cavity,blais2021circuit} or free-space optics \cite{stephenson2020high,knollmann2024integrated, chou2017note}. An alternative scenario includes the case where atoms do not emit photons, but can interact with a quantised input field~\cite{duan2004scalable}.

The different schemes considered here produce directly different classes of multipartite entangled states, such as GHZ and W states. GHZ states have many applications, including remote atomic clock comparisons \cite{komar2014quantum}, quantum secret sharing \cite{hillery1999quantum}, fundamental tests of quantum mechanics \cite{greenberger1990bell}, and teleportation \cite{chen2011spin,graham2015superdense}. They have also been proposed as a resource for distributed quantum computing \cite{gottesman1999quantum}. W states have applications in secret voting \cite{liu2011efficient}, secure quantum communication~\cite{joo2002quantum}, as a resource for teleportation \cite{shi2002teleportation,agrawal2006perfect}, metrology \cite{ng2014quantum,li2023quantum,zhang2014quantum,saleem2023achieving}, and repeaters \cite{PRXQuantum.4.040323}, among others. This article focuses on entanglement between remote nodes, but the schemes discussed here can also be employed to entangle atoms within the same node (see for example \cite{knollmann2024integrated}). In the following sections, we describe each of these schemes and how they can be extended to generate entanglement between $N$ nodes in a single shot, and find expressions for the maximum attainable entanglement fidelities and rates.

\section{Single Photon Exchange}
\subsection{Two Nodes}
Using the single photon exchange protocol, the excitation of an atom in a node can be mapped to the state of a photon. The single photon is then routed  to a different node where the inverse mapping process is applied \cite{cirac1997quantum}. If the atom in the first node is in a superposition state, then this process creates bipartite remote entanglement. This technique has been demonstrated using superconducting qubits in microwave resonators \cite{kurpiers2018deterministic,axline2018demand, campagne2018deterministic,magnard2020microwave} and with single neutral atoms in optical cavities \cite{ritter2012elementary}.

Let us consider the deterministic scheme realised using transmon-type artificial atoms in a microwave resonator, and a single microwave photon presented with number encoding, from ref.~\cite{kurpiers2018deterministic}. The three levels of the transmon, in ladder configuration, are $\ket{g}_{A_n}$, $\ket{e}_{A_n}$ and $\ket{f}_{A_n}$, with $A_n$ labeling transmon $A_1$ or $A_2$. To transfer an excitation from $A_1$ to $A_2$, first, transmon $A_1$ is prepared in the highest excited state $\ket{f}_{A_1}$, and driven coherently using the time-dependent process $\hat{g}(t)$ to the ground state $\ket{g}_{A_1}$. This process creates a microwave photon, $\ket{1}_P$ which is then coupled, with efficiency $\eta_C$, to a directional quantum channel that routes the photon to the second resonator-coupled transmon. The transmon $A_2$, initially in the ground state $\ket{g}_{A_2}$, is then driven with the time-reversed process $\hat{g} (-t)$, which has the exact opposite effect of $\hat{g} (t)$, i.e., $A_2$ absorbs a microwave photon and transfers the population from $\ket{g}_{A_2}$ to $\ket{f}_{A_2}$. Finally, the excitation transfer ends with a pulse mapping $\ket{f}_{A_2}$ to $\ket{e}_{A_2}$. 

To create entanglement between $A_1$ and $A_2$ using this approach, the following sequence is applied. First, we prepare transmon $A_1$ in a superposition state of $\ket{e}$ and $\ket{f}$, transmon $A_2$ in the ground state, and vacuum in the microwave resonators and transfer line, i.e.,
\begin{align}
    \frac{1}{\sqrt{2}}\left(\ket{e}_{A_1} + \ket{f}_{A_1}\right)\otimes\ket{g}_{A_2}\ket{0}_P,
\end{align}
where $P$ denotes the field occupation state. Then, the $\hat{g}(t)$ drive is applied in transmon $A_1$, which produces a photon only for the amplitude of the transmon that is in the state $\ket{f}$ and maps it to the ground state $\ket{g}_{A_1}$, i.e., 
\begin{align}
    \frac{1}{\sqrt{2}}\left(\ket{e}_{A_1}\ket{g}_{A_2}\ket{0}_P+ \ket{g}_{A_1}\ket{g}_{A_2}\ket{1}_P\right).
\end{align}
Then the absorption $\hat{g}(-t)$ process is applied in transmon $A_2$, producing the state $\ket{g}_{A_2}$, i.e., 
\begin{align}
    \frac{1}{\sqrt{2}}\left(\ket{e}_{A_1}\ket{g}_{A_2}\ket{0}_P+ \ket{g}_{A_1}\ket{f}_{A_2}\ket{0}_P\right).
\end{align}
Finally, a pulse transfers the amplitude from $\ket{f}_{A_2}$ to $\ket{e}_{A_2}$. Tracing out the field state, the resulting state of the transmon pair is the Bell state
\begin{align}
    \frac{1}{\sqrt{2}}\left(\ket{e}_{A_1}\ket{g}_{A_2}+ \ket{g}_{A_1}\ket{e}_{A_2}\right).
\end{align}
A weakness of this particular scheme, which uses the photon number to encode the transferred information, is that the photon state $\ket{0}_P$ can be produced either from the desired process or from photon loss, and hence, the fidelity is drastically affected by photon loss. Therefore, to achieve high fidelity, a photon transfer probability close to unity is required. In ref.~\cite{kurpiers2018deterministic}, the observed state fidelity with respect to the ideal Bell state is 78.9\%, with photon loss accounting for 10.5\% infidelity, and limited coherence times contributing 9\%. 
If the photon transmission efficiency is $\eta_{A_1A_2}$, then the  fidelity achieved by this protocol is limited to \cite{kurpiers2019quantumthesis}
\begin{align}
    F^\text{ST}_2 = \frac{1}{4}\left(1+\sqrt{\eta_{A_1A_2}}\right)^2.
\end{align}

As this is a deterministic scheme, the entanglement rate is only limited by the repetition rate of the experiment, and the time required by the photon to travel from one node to the other.

The implementation between two neutral atoms of ref.~\cite{ritter2012elementary} uses the polarisation of the emitted photon instead of the number state. For an appropriate choice of atomic structure, such a scheme can be made robust against photon loss by heralding on the successful absorption of a photon (see also ref.~\cite{piro2011heralded}). The achieved entanglement fidelity is 85\%, mostly limited by state preparation and readout, excitation of undesired electronic transitions, and multi-photon processes. By reducing the acceptance window of the herald photon, up to 98.7\% fidelity has been reached. The probability for remote entanglement creation is proportional to
\begin{align}
R^\text{ST}_{2} \propto \eta_\text{P}\eta_\text{    OUT}\eta_\text{NET}\eta_\text{ENT}\eta_\text{DET},
\end{align}
where $\eta_\text{P}$ is the probability of emitting the desired photon in the cavity mode, $\eta_\text{NET}$ is the probability of coupling the cavity mode into an optical fiber,  $\eta_\text{ENT}$ is the probability of the second atom to absorb the photon from the fiber mode, and $\eta_\text{DET}$ is the probability of detecting an emitted heralded photon. In ref.~\cite{ritter2012elementary}, entanglement is created with an overall probability of \SI{2}{\percent}. A detailed analysis of the performance of this process is presented in ref.~\cite{vogell2017deterministic}.
A similar approach has been implemented with transmon qubits and time-bin encoded photons \cite{kurpiers2019quantum}, where photon loss was detected using a non-destructive measurement of the transmon qubit state, reducing the transfer error by a factor of $\sim 2$.

\subsection{$N$ Nodes}
If we now consider the transmon scheme, but such that the generated single photon is split and routed to different nodes, it is possible to generate entanglement between $N$ different nodes. First, we prepare $N+1$ atoms in the initial state
\begin{align}
\ket{f}_{A_0}\left(\bigotimes_{n=1}^N\ket{g}_{A_n}\right).
\end{align}
Then we drive the excitation process $\hat{g}(t)$ in node 0, producing a photon $P_0$
\begin{align}
\ket{g}_{A_0}\ket{1}_{P_0}\left(\bigotimes_{n=1}^N\ket{g}_{A_n}\right).
\end{align}
Then, the photon is split, by e.g., using beam splitters \cite{gu2017microwave}, and routed to each of the nodes. If the photon is split into different paths with equal probability, the state now is
\begin{align}
    \frac{1}{\sqrt{N}}\ket{g}_{A_0}\sum\limits_{m=1}^N\ket{1}_{P_m}\left(\bigotimes_{n=1}^N\ket{g}_{A_n}\right)
\end{align}
where $\ket{1}_{P_m}$ represents the presence of a photon in resonator $m$, and no photons in all the rest, e.g., $\ket{1}_{P_2} =\ket{0}_{P_1} \ket{1}_{P_2}\ket{0}_{P_3},...,\ket{0}_{P_N}$.
If we apply the time-reversed process in all the nodes, the rotation at the end, and trace out the state of the field and atom $A_0$, then the state is the multipartite entangled W state
\begin{align}
   \ket{W_{1,N-1}}= &\frac{1}{\sqrt{N}}\left( \ket{e,g,g,...,g}_{A_1,A_2,...,A_N} \right.\nonumber\\
   &+ \left.\ket{g,e,g,...,g}_{A_1,A_2,...,A_N}\right. \nonumber\\
   & \vdots\nonumber\\
   &+ \left.\ket{g,g,g,...,e}_{A_1,A_2,...,A_N}\right).
\end{align}

If no heralded absorption is performed, i.e, the deterministic scheme is followed, then the fidelity of the generated state depends strongly on the photon losses. If the probability of transmission of the photon from atom $A_0$ to $A_N$ is the same for all the nodes and given by $\eta_{A_0,A_N}$, then the maximum attainable fidelity is proportional to
\begin{align}
    F^{\text{ST}}_N \propto \eta_{A_0,A_N}.
\end{align}
If heralded absorption is used, then the fidelity is not limited by photon loses, instead, the entanglement generation rate is limited by
\begin{align}
R^{\text{ST}}_{N}\propto \eta_\text{P}\eta_\text{OUT}\eta_\text{NET}\eta_{A_0A_N}P_\text{ENT}\eta_\text{DET},
\end{align}
where we have assumed that the different efficiencies are the same for all of the nodes. The heralding mechanism could be implemented by detecting a single photon emitted collectively by the receiver atoms, using a which-path eraser as described in section~\ref{sec:wpa}.

\section{Itinerant-Photon Controlled Entangling Gates}
\subsection{Two nodes}
Figure \ref{fig:schemes}b) illustrates a scheme in which multi-node entanglement can be achieved using an itinerant photon interacting with multiple nodes. The photon is initially prepared in a balanced superposition, and at each node, the atom interacting with the cavity is prepared in the state $\ket{0}$. The itinerant photon then interacts with each cavity through a ``reflection gate" \cite{duan2004scalable, reiserer2014quantum}. This implements a controlled-unitary interaction between the atoms and the photon. The initial state of the photon and the atoms in a two node network is given by
\begin{align}
    \frac{1}{\sqrt{2}}(\ket{0}+\ket{1})_{P}
    \ket{0}_{A_1}\ket{0}_{A_2}.
\end{align}
Here, $P$ denotes the photon, and $A_1$ and $A_2$ represent atoms in the 1st and 2nd nodes, respectively. After subsequent interaction with the first and the second cavity, and considering that the controlled gate is applied in the $X$ basis, the state of the system in the ideal and loss-less case is given by
\begin{align}           
\frac{1}{\sqrt{2}}\left(\ket{+}_{P}\ket{\Phi^+}_{A_1,A_2} + \ket{-}_{P}\ket{\Phi^-}_{A_1,A_2} \right)
\end{align}
where $\ket{\pm}=\frac{1}{\sqrt{2}}(\ket{0}\pm\ket{1})$, and the Bell state $\ket{\Phi^\pm}=\frac{1}{\sqrt{2}}\left(\ket{00}\pm\ket{11}\right)$. Subsequently, by measuring the state of the photon in the $\{\ket{\pm}\}$ basis, the states of the atoms can be projected into either Bell state $\ket{\Phi^+}$ or $\ket{\Phi^-}$. A similar proposal using continuous-variable encoding in light can be found in ref.~\cite{van2006hybrid}.

This scheme requires strong coupling between the atoms and the quantized modes of the field. It has been used to create two-node entanglement using atoms in optical cavities and photons with polarisation encoding \cite{welte2018photon,langenfeld2021quantum}, and with SiV-centers in photonic crystal cavities and photons with time-bin encoding \cite{knaut2024entanglement}. In both cases, the achieved entanglement fidelities are below 90\%, mostly limited by the fidelity of the photon-atom gates and the use of weak coherent photon pulses instead of single photons. The photon-atom gate fidelities themselves are limited by spatial mismatch and the finite cooperativity of the atom-cavity systems. The entanglement rate is limited by the experiment repetition rate, optical losses, and the finite cooperativity. In both experimental platforms, a strong trade-off between the fidelity and the average number of photons in the coherent pulse was observed \cite{welte2018photon,langenfeld2021quantum, knaut2024entanglement}, resulting, in practice, in a trade-off between fidelity and entanglement rate. The best observed entangling rate in the neutral atom experiment is \SI{60}{\per\second} \cite{langenfeld2021quantum}, while in the SiV-centre experiment, it is \SI{0.2}{\per\second} \cite{knaut2024entanglement}, both corresponding to fidelities of $\sim 2/3$.

A detailed study of the interplay between photon-atom gate fidelity versus cooperativity of the atom-cavity system is presented in ref.~\cite{wang2016universal}. Details of how the mismatch between the input photon and the cavity affect the fidelities can be found in ref.~\cite{reiserer2014quantum}.

Assuming that both nodes can perform photon-controlled gates on the atoms, leading to the same atom-photon entangled fidelity $F_{\text{A,P}}$; if the error of the controlled gate can be fully described by a depolarising channel, the fidelity of the generated atom-atom entangled state will be limited by
\begin{align}
    F^\text{I}_2  = 2F_{\text{P,A}}-1,
\end{align}
where $F_{\text{A,P}}$ is the fidelity of the atom-photon system. 

\subsection{$N$ Nodes}
\begin{figure}
\centerline{\includegraphics[width=0.8\columnwidth]{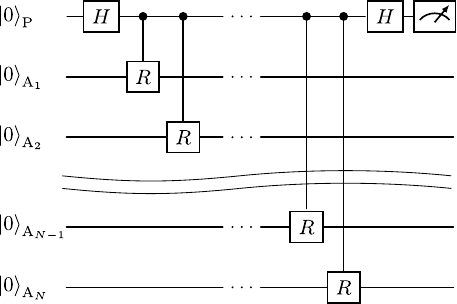}}
\caption{Quantum circuit for $N$-node entanglement using the Itinerant Photon scheme. $R$ is controlled rotation. }
\label{fig:itinerant_circuit}
\end{figure}

The extension of this protocol to an $N$-node network is depicted in Fig \ref{fig:itinerant_circuit}. The initial state of the photon and the $N$
nodes is 
\begin{align}
    \ket{+}_{P} \bigotimes\limits_{n=1}^{N}
    \ket{0}_{A_n} ,
\end{align}
where $A_n$ denotes the state of an atom in the $n^\text{th}$ cavity. After the interaction with each cavity, and considering controlled gates in the $X$ basis, the state of the system in the ideal, loss-less case is
\begin{align}
\frac{1}{\sqrt{2}}\left(\ket{+}_{P}\ket{G_0^+}_{A_1,A_2,..,A_n} + \ket{-}_{P}\ket{G_0^-}_{A_1,A_2,..,A_n} \right)
\end{align}
with the $N$-particle GHZ states
\begin{align}
\ket{G_0^\pm}=\frac{1}{\sqrt{2}} \left(\ket{00...0}\pm\ket{11...1}\right) .
\end{align}
Thereafter, measuring the state of the photon in the $\{\ket{\pm}\}$ basis projects the states of the atoms either into the multipartite entangled state $\ket{G_0^+}$ or into $\ket{G_0^-}$.

\begin{figure}
\centerline{\includegraphics[width=1.0\columnwidth]{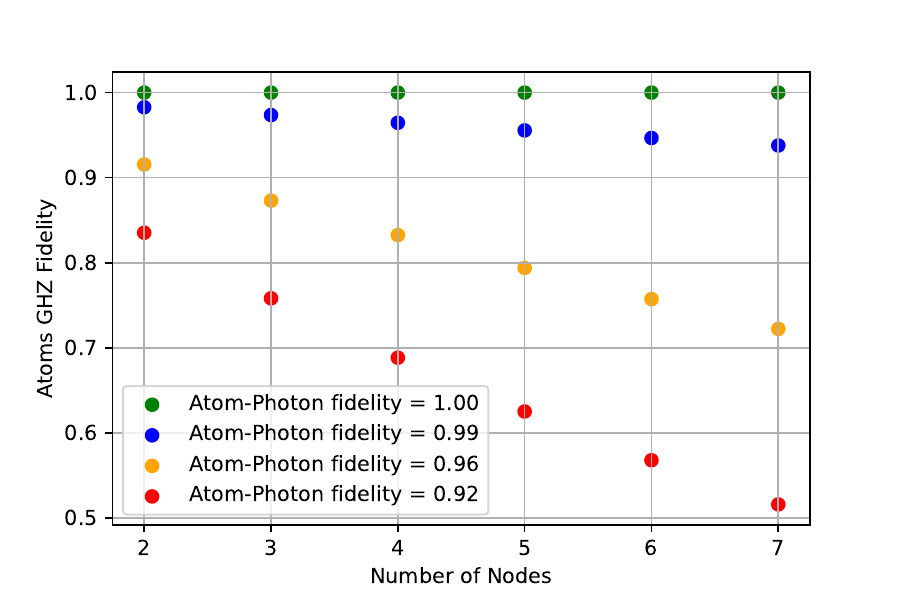}}
\caption{ Fidelities for the GHZ states generated across several nodes  using the itinerant photon method. It is assumed that the only imperfect process is the controlled gate between the photon and atom in each atom, achieving a limited Bell state fidelity, which is identical for all of the nodes. }
\label{fig:itinerant_several_nodes}
\end{figure}

If the dominant error mechanisms are the same as in the 2-node implementations, i.e., the fidelity of the photon-controlled atom gates,  the fidelity of the resulting multipartite entangled state decreases linearly with increasing number of nodes. : Fig.~\ref{fig:itinerant_several_nodes} shows how the maximum attainable fidelity depends on the number of nodes and the atom-photon entangled state fidelity, where the controlled gates is performed with the same fidelity in all nodes. Importantly, as the success is heralded by the detection of a single photon, photon losses do not affect the atom state fidelity.
The rate of entanglement generation depends on the transmission efficiency of the photon along the different nodes. If the probability of transmission from one to another is $\eta_\text{T}$, the probability of the photon not being lost during the interaction with that atom-cavity system is  $\eta_\text{C}$, and $\eta_\text{DET}$ is the efficiency of the photon detection at the end, then the success probability is proportional to
\begin{align}
    P_N \propto \eta_\text{T}^{N-1}\eta_\text{C}^N\eta_\text{DET}.
\end{align}

Using the fidelities and rates demonstrated in refs.~\cite{langenfeld2021quantum,knaut2024entanglement} it would be possible to create GHZ states across several nodes, although with low fidelity and rates. New schemes for lower-loss and higher-fidelity atom-photon gates have been proposed \cite{nagib2023robust}, which, together with higher cooperativity resonators and better single photon sources could achieve better performance for this networking scheme.

\section{Mapping Entanglement from Photons to Atoms}
\subsection{Two nodes}
\label{sec:mapping}
\begin{figure}
\centerline{\includegraphics[width=1.0\columnwidth]{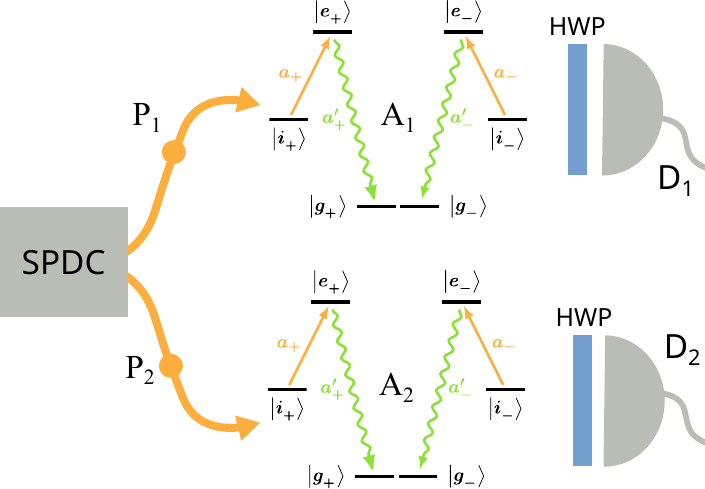}}
\caption{The photon-to-atom entanglement mapping scheme for two nodes. A SPDC source is used to generate a polarisation-entangled pair of photons ($P_1$ and $P_2$), which are probabilistically absorbed by two atoms ($A_1$ and $A_2$) with double-$\Lambda$ electronic structure. Upon successful absorption, a photon is emitted from the atoms, and the photons are rotated using a half-wave plate (HWP), providing indistinguishability between the absorption channels. The detection of a photon from each of the atoms heralds the successful creation of entanglement between the atoms.}
\label{fig:photon_scheme}
\end{figure}

Fig.~\ref{fig:schemes}c depicts a scheme where a multi-photon entangled source (MPES) is used to generate entanglement across different nodes. Let us first consider the case of two nodes. Following ref.~\cite{sangouard2013heralded}, a photon pair source produces polarisation-entangled pairs, for instance, photons in the state
\begin{align}
    \frac{1}{\sqrt{2}}\left(\ket{a_+}_{P_1}\ket{a_-}_{P_2}-\ket{a_-}_{P_1}\ket{a_+}_{P_2}\right)
\end{align}
where $P_1$ and $P_2$ denote the two photons, and $\{\ket{a_\pm}\}$ are two orthogonal polarisation states. Let us also consider two separated network nodes, each with an atom with double-$\Lambda$ level structure, see Fig.~\ref{fig:photon_scheme}, initially prepared in a separable superposition state
\begin{align}
    \frac{1}{\sqrt{2}}\left( \ket{i_+}_{A_1}+\ket{i_-}_{A_1}\right)\otimes\frac{1}{\sqrt{2}}\left( \ket{i_+}_{A_2}+\ket{i_-}_{A_2}\right).\label{eq:atoms_EPR_initial}.
\end{align}
The absorption of an $a_+$-polarised photon excites the atomic Raman transition $\ket{i_+}\rightarrow\ket{e_+}\rightarrow \ket{g_+}$ emitting an $a^\prime_+$-polarised photon, whereas the absorption of an $a_-$-polarised photon excites the atomic Raman transition $\ket{i_-}\rightarrow\ket{e_-}\rightarrow \ket{g_-}$, emitting an $a^\prime_-$-polarised photon. If each photon is sent to a different atom, upon absorption and emission of a photon, the state of the atoms and the emitted field is
\begin{align}
\frac{\left(\ket{g_+}_{A_1}\ket{g_-}_{A_2}\ket{a^\prime_{+}}_{1}\ket{a^\prime_{-}}_{2} - \ket{g_-}_{A_1}\ket{g_+}_{A_2}\ket{a^\prime_{-}}_{1}\ket{a^\prime_{+}}_{2}\right)}{\sqrt{2}},
\end{align}
where $\ket{a^\prime_{\pm}}_{n}$ denotes the emission of a heralding photon with polarisation $a^\prime_\pm$ from atom $n=1\text{ or }2$. The detection of the emitted photons from each of the atoms in a basis that makes the orthogonal polarisations indistinguishable, e.g., $\{\frac{1}{\sqrt{2}}(\ket{a^\prime_+}+\ket{a^\prime_-} ),\frac{1}{\sqrt{2}}(\ket{a^\prime_+}-\ket{a^\prime_-} )\}$, projects the state of the atoms into a Bell state
\begin{align}
    \ket{\Psi^{\pm}}=\frac{1}{\sqrt{2}}\left(\ket{g_+,g_-}_{A_1A_2}\pm\ket{g_-,g_+}_{A_1A_2}\right),
\end{align}
where the sign depends on the parity of the photon measurement outcomes.
If we consider a pulsed entangled photon source which produces a polarisation-entangled pair with probability $p_\text{EPR}$, where the probability of a photon being absorbed by one of the atoms is $\eta_\text{ABS}$, and the probability of a heralding photon from one of the atoms being emitted and detected is $\eta_\text{DET}$, the total probability of creating a two-atom entangled state is
\begin{align}\label{P-EM-2}
    P^\text{EM}_2 \propto p_\text{EPR}\left(\frac{1}{2}\eta_\text{ABS}\eta_\text{DET}\right)^2\,
\end{align}
where the $\frac{1}{2}$ factor comes from the fact that the $\ket{i_+}_{A_1}\ket{i_-}_{A_2}$ and $\ket{i_-}_{A_1}\ket{i_+}_{A_2}$ terms in the initial state of the atoms [Eq.~\eqref{eq:atoms_EPR_initial}] do not contribute to the absorption process.
An important issue of this scheme is that the photons from the entangled source need to match the linewidth, frequency, and wavepacket of the time-inverted atomic emission process. This can be done partially by using a filter cavity at the price of a considerable loss of photons \cite{schuck2010resonant}.
Using free-space optics to couple the single photons onto the atoms, Shuck et al.~\cite{schuck2010resonant} achieved $\eta_\text{ABS}\sim\num{e-4}$ and a collection efficiency of the heralded photons of $\eta_\text{DET}\sim\SI{5}{\percent}$. A pulsed photon such as the one of ref.~\cite{jons2017bright} can produce entangled pairs with probability $p_\text{EPR}\sim\num{2.5e-3}$ per pulse. This adds up to a total probability of $P_2\sim\num{e-13}$. Using an optical resonator to couple the atoms could increase the probabilities substantially, up to $\sim\num{e-5}$, using the cavity parameters discussed in ref.~\cite{pfister2016quantum}. 

The achievable fidelities are limited by the fidelity and purity of the entangled photon source, $F_2^\text{ph}$, and given the extremely low rates of heralding photons, by false heralds due to dark counts in the detectors. The fidelity is therefore limited to
\begin{align}
F^\text{EM}_2 =  \frac{F_2^\text{ph} P_2^\text{EM}+\frac{1}{4} P_2^\text{false}}{P_2^\text{EM}+P_2^\text{false}},
\end{align}
assuming a maximally mixed state when a false herald occurs with probability $P_2^\text{false}$.
The effect of false heralds can be lowered by verifying that there is no population left in $\{\ket{i_{+}},\ket{i_{-}}\}$ by a subspace measurement on the atoms, to herald whether the atom absorbed the photon. Such a measurement would keep the coherence within the target qubit subspace $\{\ket{g_{+}},\ket{g_{-}}\}$ and otherwise flag unsuccessful attempts with high fidelity. The dark count probability then only competes with the local detection efficiency rather than with the overall efficiency $P_2^\text{EM}$.

\subsection{$N$ Nodes}
To scale-up this scheme and generate entanglement between $N$ nodes simultaneously, we require a multi-photon entangled source (MPES). For instance, let us consider that the atoms are prepared in the separable superposition state
\begin{align}
    \bigotimes\limits_{n=1}^N\frac{1}{\sqrt{2}}\left(\ket{i_+} + \ket{i_-}\right)_{A_n}.
\end{align}
If the source produces a polarisation-encoded $N$-photon GHZ state \cite{zhong201812, thomas2022efficient,cao2023photonic}
\begin{align}
\ket{G_0^+}_{P_1,P_2,...,P_N} = \frac{1}{\sqrt{2}}\left(\ket{a_+,a_+,...,a_+} +\ket{a_-,a_-,...,a_-} \right),    
\end{align}
then, upon absorption of the source photons by the atoms in the nodes, and subsequent emission and detection of $N$ herald photons, the state of the atoms becomes
\begin{align}
\ket{G_0^{\pm}}_{A_1,A_2,,..,A_N} = \frac{1}{\sqrt{2}}\left(\ket{g_+,g_+,...,g_+} \pm \ket{g_-,g_-,...,g_-} \right),
\end{align}
where the sign depends on the parity of the photon measurement outcomes.
The successful completion of this process requires coincident detection of $N$ photons, with probability proportional to
\begin{align}
    P^\text{EM}_N \propto p_{\text{GHZ},N}\left(\frac{1}{2}\eta_{\text{ABS}}\eta_{\text{DET}}\right)^N.
\end{align}
Again, the main limitation for the fidelities here will be the initial entangled multi-photon fidelity, and given the extremely low process rates, the dark counts in the detectors. If each of the heralding detectors has a probability $p_\text{dark}$ of measuring a dark count during the detection window, then the probability for a false herald is
\begin{align}
    P_N^\text{false} = \sum\limits_{n=0}^N \binom{N}{n}p_\text{real}^n p_\text{dark}^{N-n} ,
\end{align}
which represents the fact that the simultaneous detection of $N$ heralding photons can be triggered by any combination of detections of $n$ ``real" heralding photons and $N-n$ ``dark" photons, with $n\in[0,N]$. Thus, the fidelity of the generated state is limited by
\begin{align}
    F^\text{EM}_N = \frac{F_N^\text{ph}  p_N^\text{EM} + \frac{1}{2^N} P_N^\text{false}}{p_N^\text{EM}  + P_N^\text{false}} .
\end{align}

Importantly, this $N$-node scheme is not restricted to mapping GHZ states from photons to atoms, but, in principle, any multi-photon state could be mapped.

\section{Remote Entanglement via Entanglement Swapping}
\label{sec:swap}
\subsection{Two Nodes}
Fig.~\ref{fig:schemes}d illustrates the scheme for remote entanglement via entanglement swapping from atom-photon pairs to atom pairs by performing a Bell measurement on the states of the photons, as proposed in refs.~\cite{feng2003entangling, duan2003efficient, simon2003robust}.

Let us start by considering the standard two-atom case. The four Bell states are defined as
\begin{align}
\label{Bell_states}
\left|\Phi^{\pm}\right\rangle = \frac{1}{\sqrt{2}}\left(\left|00\right\rangle \pm \left|11\right\rangle\right), \quad
\left|\Psi^{\pm}\right\rangle = \frac{1}{\sqrt{2}}\left(\left|01\right\rangle \pm \left|10\right\rangle\right).
\end{align}
One can then express the standard computational basis states in terms of the Bell states:
\begin{align}
&\left|00\right\rangle = \frac{1}{\sqrt{2}}\left(\left|\Phi^{+}\right\rangle + \left|\Phi^{-}\right\rangle\right), 
\quad
\left|11\right\rangle = \frac{1}{\sqrt{2}}\left(\left|\Phi^{+}\right\rangle - \left|\Phi^{-}\right\rangle\right), \notag\\
&\left|01\right\rangle = \frac{1}{\sqrt{2}}\left(\left|\Psi^{+}\right\rangle + \left|\Psi^{-}\right\rangle\right), 
\quad
\left|10\right\rangle = \frac{1}{\sqrt{2}}\left(\left|\Psi^{+}\right\rangle - \left|\Psi^{-}\right\rangle\right).
\end{align}
The entangled state
\begin{align}
    \frac{1}{\sqrt{2}}\left(\left|00\right\rangle \pm \left|11\right\rangle\right)_{A_1P_1} \otimes
	\frac{1}{\sqrt{2}}\left(\left|00\right\rangle \pm \left|11\right\rangle\right)_{A_2P_2}
\end{align}
of two atoms ($A_1$ and $A_2$) and two photons ($P_1$ and $P_2$) can then be expressed as
\begin{align}
   &\phantom\pm \frac{1}{2}\left|\Phi^{+}\right\rangle_{A_1A_2} \left|\Phi^{+}\right\rangle_{P_1P_2} +
	\frac{1}{2}\left|\Phi^{-}\right\rangle_{A_1A_2} \left|\Phi^{-}\right\rangle_{P_1P_2}\nonumber\\
 &\pm 	\frac{1}{2}\left|\Psi^{+}\right\rangle_{A_1B_2} \left|\Psi^{+}\right\rangle_{P_1P_2} \pm
	\frac{1}{2}\left|\Psi^{-}\right\rangle_{A_1B_2} \left|\Psi^{-}\right\rangle_{P_1P_2}. 
\end{align}

\begin{figure}
\centerline{\includegraphics[width=0.8\columnwidth]{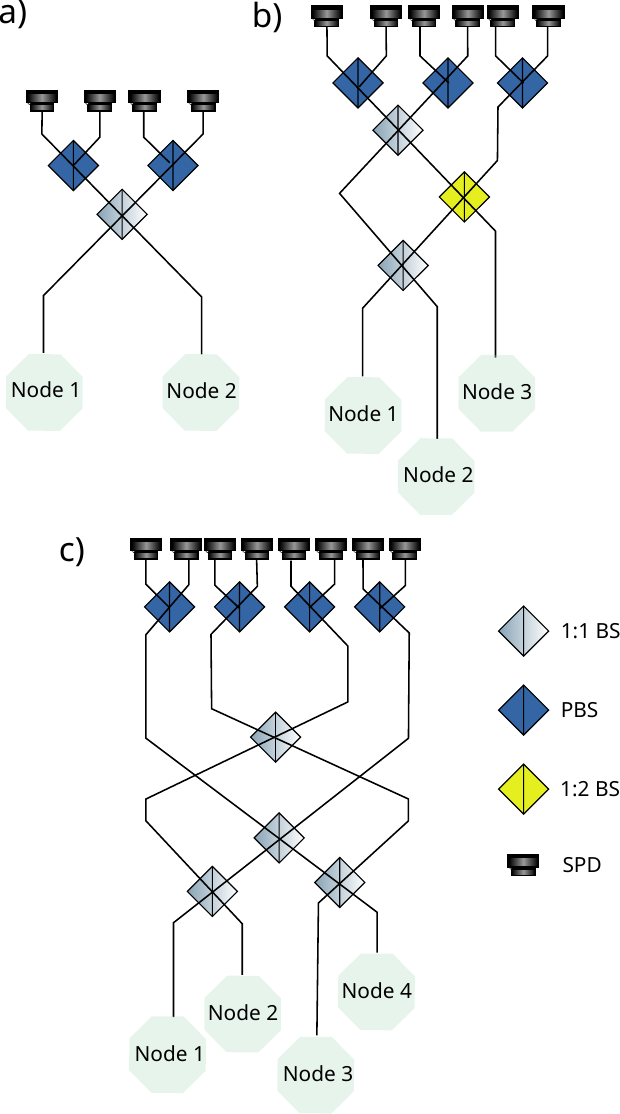}}
\caption{a) Standard polarisation partial Bell state analyser for 2-node entanglement. b) Generalised Bell state analyser for 3-node entanglement. c) Generalised Bell state analyser for 4-node entanglement. }
\label{fig:GBSA_2_3_4}
\end{figure}

 For the photons, different degrees of freedom have been proposed and experimentally demonstrated, including polarisation, time-bin, and energy. For example, if the photon degree of freedom used is polarisation, with $\ket{H}$ and $\ket{V}$ representing horizontal and vertical polarisations respectively, we use the mapping
\begin{align}
&\ket{0}\leftrightarrow\ket{H}, \quad
    \ket{1}\leftrightarrow\ket{V}.
\end{align}
A partial Bell state measurement on the photon polarisation state can be implemented using the setup in Fig.~\ref{fig:GBSA_2_3_4}a. Upon the coincident detection of two clicks on different detectors, the outcome of the photons' Bell measurement is either $\ket{\Psi^+}$ or $\ket{\Psi^-}$. Therefore, these coincident clicks herald the creation of entanglement between the two atoms via entanglement swapping. Double detections in any single detector correspond to non-entangled states. Note that this limits the entanglement generation success probability to 1/2, which is the maximum efficiency allowed for a Bell state analyser (BSA) that uses only linear optics \cite{calsamiglia2001maximum}. Importantly, unlike the cases shown below for more particles, the efficiency of 1/2 cannot be improved by using number-resolved detectors. A similar Bell-state analyser for time-bin encoded photons was proposed in \cite{valivarthi2014efficient}, which also achieves the theoretical maximum efficiency.

If the probability of detecting a photon emitted by two different nodes, $A$ and $B$, is equal, and given by $\eta_{\text{DET}}$, then the probability of creating entanglement between the two atoms heralded by the detection of two photons in coincidence is
\begin{align}
    R^\text{SW}_2 \propto \frac{1}{2}\eta_{\text{DET}}^2.
\end{align}

Importantly, in order to achieve high fidelities, the photons need to be completely indistinguishable. This includes the temporal wavepacket of the photons, necessitating the synchronisation of the arrival times of the photons at the BSA. 
So far, the fidelity of the created states has been limited by photon distinguishability, the atom-photon entanglement, and by imperfect optical systems. The atom-photon state fidelities are limited by the photon collection geometry, the electric field emission patterns of the involved transitions, and the purity of the emitted single photon (see refs.~\cite{stephenson2020high, krutyanskiy2023entanglement, seubert2020analyzing} for details).

This scheme has been implemented with neutral atom ensembles \cite{chou2005measurement, chou2007functional, bao2012quantum}, trapped ions, using photon-polarisation encoding and free-space optical collection \cite{moehring2007entanglement, stephenson2020high}, trapped-ions using photon time-bin encoding and free space optical collection \cite{saha2024high}, trapped ions coupled to optical cavities using photon polarisation encoding \cite{krutyanskiy2023entanglement}, single neutral atoms and free-space optical collection \cite{hofmann2012heralded}, and NV-centres and time-bin encoded photons \cite{barrett2005efficient,hensen2015loophole}.
The best remote Bell-state fidelity achieved using this scheme was \SI{96}{\percent} at a rate of $\sim 100$ entangled events per second \cite{nadlinger2022experimental}. For experiments based on photons emitted by spontaneous decay, the final atom-atom entanglement fidelity is independent of the generation rate \cite{stephenson2020high}. On the other hand, experiments based on photons created via stimulated emission do suffer from a strong trade-off between maximum achievable fidelities and rates \cite{tanji2024rate}.

\subsection{Three Nodes}
Next, consider the same scenario with three identical atom-photon pairs. We denote the atoms as \( A_1 \), \( A_2 \), and \( A_3 \), and the photons as \( P_1 \), \( P_2 \), and \( P_3 \). The state can be described by
\begin{align}
\label{ES_three_identical_Bell1}
&\bigotimes\limits_{n=1}^3 \frac{1}{\sqrt{2}}\left( \ket{0}_{A_n}\ket{0}_{P_n}\pm\ket{1}_{A_n}\ket{1}_{P_n}\right)=\nonumber\\
{} &	\frac{1}{\sqrt{8}} \big(\left|G_0^{+}\right\rangle_{A_1A_2A_3}\left|G_0^{+}\right\rangle_{P_1P_2P_3} +
	\left|G_0^{-}\right\rangle_{A_1A_2A_3} \left|G_0^{-}\right\rangle_{P_1P_2P_3} \nonumber\\
 &+ 	\left|G_1^{+}\right\rangle_{A_1A_2A_3}  \left|G_1^{+}\right\rangle_{P_1P_2P_3}  +
	\left|G_1^{-}\right\rangle_{A_1A_2A_3}  \left|G_1^{-}\right\rangle_{P_1P_2P_3} \nonumber \\
 &+ 	\left|G_2^{+}\right\rangle_{A_1A_2A_3}  \left|G_2^{+}\right\rangle_{P_1P_2P_3}  +
	\left|G_2^{-}\right\rangle_{A_1A_2A_3}  \left|G_2^{-}\right\rangle_{P_1P_2P_3}\nonumber\\
 &+	
	\left|G_3^{+}\right\rangle_{A_1A_2A_3}  \left|G_3^{+}\right\rangle_{P_1P_2P_3}  +
	\left|G_3^{-}\right\rangle_{A_1A_2A_3}  \left|G_3^{-}\right\rangle_{P_1P_2P_3}  \big) .
\end{align}
The $\{\ket{G_n^\pm}\}$ states are GHZ states that form a complete orthonormal basis in the 3-qubit Hilbert space. Explicitly, they are given by
\begin{align}
    \ket{G_0^{\pm}} &= \frac{1}{\sqrt{2}}\big(\ket{000} \pm \ket{111} \big)\ ,\\
    \ket{G_1^{\pm}} &= \frac{1}{\sqrt{2}}\big(\ket{001} \pm \ket{110} \big)\ ,\\
    \ket{G_2^{\pm}} &= \frac{1}{\sqrt{2}}\big(\ket{010} \pm \ket{101} \big)\ ,\\
     \ket{G_3^{\pm}} &= \frac{1}{\sqrt{2}}\big(\ket{011} \pm \ket{100} \big)\ .
\end{align}
The challenge is to implement a 3-photon generalized Bell state analyser (BSA) that can distinguish between different photonic $\ket{G}$ states and project the corresponding 3-particle entangled state in the atoms, performing the entanglement swapping operation. Following the idea used for the 2-node BSA, such a generalised GHZ Bell state analyser requires the coincident detection of 3 photons and the capability of erasing any information about which atom emitted which photon. This problem has been considered before in photonic networks for the creation of multipartite photon entanglement using interferometers in post-selection experiments (see, e.g., refs.~\cite{pan1998greenberger, lougovski2009verifying}. For the case of entangling three atoms, the requirement can be satisfied by a 3-to-3 symmetric multiport \cite{zukowski1997realizable,lim2005multiphoton}, which is a generalisation of a 50:50 beam splitter to 3 inputs and 3 outputs. In a 3-to-3 symmetric multiport, a photon entering any port has the same probability of exiting any of the 3 output ports. An example of a 3-to-3 symmetric multiport is the tritter \cite{campos2000three}. Fig.~\ref{fig:GBSA_2_3_4}b shows a generalised 3-photon BSA implemented using a tritter, which can also be constructed using integrated photonics \cite{spagnolo2013three}. The tritter uses two 50:50 beam splitters and one 1:2 beam splitter (transmission 2/3 and reflection 1/3).
The creation of 3-particle entanglement is heralded by the detection of 3 photons in any combination of three different detectors.

Different implementations of the 3-photon generalised BSA will lead to different probabilities of generating entanglement. We label this probability as $p_{\text{BSA}_3}$. The success probability of this 3-photon heralded scheme to generate 3-particle entanglement is thereafter
\begin{align}
   R^\text{SW}_3 \propto p_{\text{BSA}_3}\eta_{\text{DET}}^3,
\end{align}
where it is assumed that  photons from different nodes have the same probability of being detected ($\eta_\text{DET}$).

For the particular implementation shown in Fig.~\ref{fig:GBSA_2_3_4}b, using photon polarisation encoding, the probability of generating 3-particle entanglement by heralding in three different detectors and using single-photon detectors is $p_{\text{BSA}_3}=1/4$. If number-resolved photodetectors are available, the probability of generating a 3-particle entangled state heralded by the detection of three photons is $3/4$. Notice that this scheme generates states of the GHZ class and of the W class, depending on the detection pattern (see appendix B).
A different proposed generalised BSA can be found in ref.~\cite{vivoli2019high}, achieving 3-particle entanglement with probability proportional to $p_{\text{BSA}_3}$= 1/8. 
In both cases, an interferometer is used to entangle atoms using photons with polarisation, time-bin \cite{valivarthi2014efficient}, or energy encoding \cite{connell2021ion}. Other proposed interferometers can be found in ref.~\cite{tanji2024rate}.

\subsection{$N$ Nodes}
\label{sec:swap_N}
Let us now consider the case of $N$ pairs of atom-photon entangled states, i.e.,
\begin{align}
\bigotimes \limits_{n=1}^{N} \frac{1}{\sqrt{2}} \big( \ket{00}_{A_n P_n} + \ket{11}_{A_n P_n} \big),   
\end{align}
where $A_n$ denotes an atom and $P_n$ a photon. This state can be written as \cite{ji2022entanglement}
\begin{align}
	\frac{1}{\sqrt{2^N}}\sum\limits_{n=0}^{2^{N-1}} \ket{G_n^+}_{A_1 \dots A_N}\ket{G_n^+}_{P_1 \dots P_N} + \ket{G_n^-}_{A_1 \dots A_N}\ket{G_n^-}_{P_1 \dots P_N}.
\label{eq:gen_N}
\end{align}
This equation shows that $N$ entangled atom-photon pairs can be written as a superposition of states where the atom and photon part are separated, but each part is entangled.
The states $\ket{G_n^\pm}$ are defined by
\begin{align}
\left|G_{n}^{\pm}\right\rangle = \frac 1{\sqrt{2}}
\Big( \left|B\big(n\big)\right\rangle \pm 
			\left|B\big(2^{N}-n-1 \big)\right\rangle \Big),
\end{align}
where $n = 0,1,\ldots,2^{N-1}-1$, and $B(n) = 0b_2b_3\cdots b_N$ is the
binary representation of $n$ in an bit string of length $N$, and thereafter
$n = \sum_{k=2}^{N} b_k\cdot2^{N-k}$.
The states ${\ket{G_n^\pm}}$ are a set of $2^N$ elements, and form an orthonormal basis for the $N$ qubits Hilbert space, i.e., $\langle G_{n}^{\pm}|G_{n^{\prime}}^{\pm} \rangle = \delta_{n,n^{\prime}}$. The computation basis elements $\ket{B(n)}$ are related to the $\ket{G_n^\pm}$ states by
\begin{align}
& \phantom{i} \left|B\big(n\big)\right\rangle = \frac 1{\sqrt{2}}
\Big( \left|G_{n}^{+}\right\rangle + \left|G_{n}^{-}\right\rangle \Big), \\
& \phantom{i} \left|B\big(2^{N}-n-1 \big)\right\rangle = \frac 1{\sqrt{2}}
\Big( \left|G_{n}^{+}\right\rangle - \left|G_{n}^{-}\right\rangle \Big).
\end{align}
The $\ket{G_n^\pm}$ states can also be written as
\begin{align}
\left|G_{n}^{\pm}\right\rangle = \frac 1{\sqrt{2}}
\Big( \left|0\right\rangle \bigotimes_{k=2}^{N} \left|b_k\right\rangle \pm
\left|1\right\rangle \bigotimes_{k=2}^{N}  \left|\bar{b}_k\right\rangle \Big).
\end{align}

By projecting the $N$-photon state to any of the $\ket{G_n^{\pm}}$ states, or superpositions, we project the state of the atoms to the corresponding $N$-particle state. As before, if an $N$-photon generalised BSA can be constructed, and the entanglement can be heralded by the detection of $N$ photons yielding to an entangled state with probability $p_{\text{BSA},N}$, then the probability of generating $N$-particle entanglement is given by 
\begin{align}
    P_{\text{BSA}_N}\eta^N.
\end{align}

An explicit calculation for $N=4$ is shown in appendixes \ref{app:Swapp_4_nodes} and \ref{app:quarters}.  
Thereafter, the challenge is to build a generalised BSA that can herald on the detection of $N$-photons and generate entanglement with some probability. For this purpose, it is important that the generalised BSA erases the which-input information. This can be achieved using an $N$-to-$N$ symmetric multiport \cite{zukowski1997realizable,lim2005multiphoton}. In an $N$-to-$N$ symmetric multiport, a single photon entering any of the $N$ inputs has the same probability of exiting any of the outputs, i.e., probability $1/N$ \cite{zukowski1997realizable}. 
For $N = 2^d$, with integer $d$, a generalised BSA can be constructed by using a $2^d$-to-$2^d$ symmetric mutiport composed of multiple layers of 50:50 beams splitters, such that each input mode is mixed with every other input mode. Each of the $2^d$ input photons crosses $d$ 50:50 beam splitters, and has a probability of $\frac{1}{2^d}$ of exiting via any of the outputs. Fig.~\ref{fig:GBSA_2_3_4}b shows a particular implementation for the 4-node case, and in appendix \ref{app:quarters} we explicitly calculate the different states produced by each possible detection pattern. In appendix \ref{app:16Node} we show how to construct a general $2^d$-to-$2^d$ symmetric mutiport. 

Importantly, a $2^d$-to-$2^d$ symmetric mutiport allows for entanglement between any sub-network of $m$ = 2, 3,...,  $2^d-1$ nodes. For example, we can entangle 3 nodes by emitting a photon from each of them using a 8-to-8 interferometer, heralded by the detection of 3 photons in any combination of output ports. This is beneficial, as it does not require the use of additional optical switches.

Alternative generalised BSAs have been considered in refs.~\cite{wang2009schemes,vivoli2019high}.

\section{Remote Entanglement via Which-Path Erasing}
\label{sec:wpa}
\subsection{Two Nodes}
The ``which-path erasing'' scheme, also known as the \textit{Cabrillo} or photon-number scheme, is based on the detection of a single photon emitted indistinguishably by one of the two atoms \cite{cabrillo1999creation,slodivcka2013atom,araneda2018interference,humphreys2018deterministic}. An excitation scheme and appropriate levels are chosen such that the photon emission is the indicator of the atom being in a particular state. 
If we are able to detect a single photon, but the information of which atom emitted the photon has been erased, the measurement of that photon projects the atoms into a multipartite entangled state.

Let us consider that two atoms ($n=1,2$) are initially prepared in a superposition state
\begin{align}
    \sqrt{1-p_n}\ket{g_-}_{A_n} + \sqrt{p_n}\ket{g_+}_{A_n}.
\end{align}
Then, a $\pi$ pulse transfers the population from $\ket{g_+}$ to a fast-decaying excited state $\ket{e}$, which decays back to the $\ket{g_+}$ state by emitting a photon. Provided that the excitation pulse is much faster than the decay rate of the excited state, the state of one atom is described by
\begin{align}
    \sqrt{1-p_n}\ket{g_-,0}_{A_n,P_n} + e^{i\phi_{D,n}+i\phi_{L,n}}\sqrt{p_n}\ket{g_+,1}_{A_n,P_n},
\end{align}
where $\{\ket{0}, \ket{1}\}$ is the photon occupation number of the emitted field, $\phi_{L,j}$ is the phase of the laser driving the transition $\ket{g_+}\rightarrow\ket{e}$ at the position of atom $n$, and $\phi_{D,n}$ is the phase acquired by the photon emitted by atom $j$ on its path to the detector.

If two atoms simultaneously go through this process, the total state of the system is given by
\begin{align}
     & \sqrt{1-p_1}\sqrt{1-p_2}\ket{g_-,g,0,0}\nonumber\\
    & + \sqrt{p_1}\sqrt{1-p_2} e^{i(\phi_{L,1}+\phi_{D,1})}\ket{g_+,g_-,1,0}\nonumber\\
     & + \sqrt{1-p_1}\sqrt{p_2} e^{i(\phi_{L,2}+\phi_{D,2})}\ket{g_-,g_+,0,1}\nonumber\\
     & + \sqrt{p_1}\sqrt{p_2} e^{i(\phi_{L,1} + \phi_{L,2} + \phi_{D,1} + \phi_{D,2}  )} \ket{g_+,g_+,1,1}.
\end{align}
\begin{figure}
\centerline{\includegraphics[width=0.8\columnwidth]{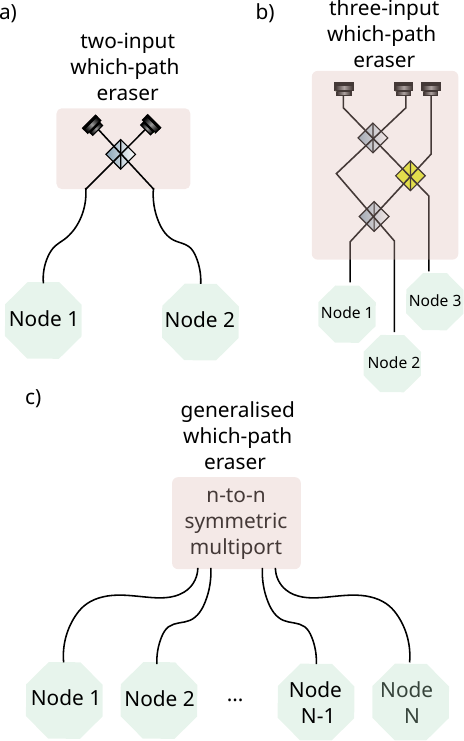}}
\caption{Photon-number based remote entanglement for a) two nodes using a beam splitter, b) three nodes using a tritter, and c) using an $N$-to-$N$ symmetric multiport for $N$ nodes.}
\label{fig:cabrillo}
\end{figure}
Then, we direct the collected photons to a which-path eraser (WPE), as shown in Fig.~\ref{fig:cabrillo}a. For the case of two inputs, this can be implemented using a 50:50 beam splitter. If the detected modes of the emission from the atoms overlap perfectly, i.e., the photons emitted from either atom are indistinguishable, the detection of a single photon in either detector projects the state of the atoms into the (unnormalised) state 
\begin{align}
    & \sqrt{p_1}\sqrt{1-p_2} e^{i(\phi_{L,1}+\phi_{D,1})}\ket{g_+,g_-}\nonumber\\
   & + \sqrt{1-p_1}\sqrt{p_2} e^{i(\phi_{L,2}+\phi_{D,2})}\ket{g_-,g_+}.
\end{align}
 This will occur with probability $\eta_1 p_1(1-p_2)+\eta_2 p_2(1-p_1)$, where $\eta_{1,2}$ is the overall detection efficiency of the emitted single photon from atom 1 and atom 2, respectively.

 If the probability of excitation is the same for both atoms, $p_1 = p_2 = p$, and the efficiency of the collection, transmission, and detection is the same for both, i.e., $\eta_1 = \eta_2 = \eta$, the generated state is approximately given by
 \begin{align}
     \ket{\Psi^\phi} = \frac{1}{\sqrt{2}}(\ket{g_+,g_-}\pm e^{i\phi}\ket{g_-,g_+}),
 \end{align}
 with an overall probability of $2\eta p(1-p)$ and $\phi=(\phi_{L,2}-\phi_{L,1}) + (\phi_{D,B}-\phi_{D,A})$. The sign of the superposition depends on what output of the beam splitter the photon was detected.

Notice that there is a non-vanishing probability,  proportional to $p^2$, of exciting both atoms and detecting just one photon. If $p$ is kept small, then $p^2$ is negligible. Thereafter, upon a click in one of the detectors, the state is 
\begin{align}
   & \frac{1}{\sqrt{2p(1-p)+p^2}}\times\nonumber\\
    &\left[\frac{\sqrt{2p(1-p)}}{\sqrt{2}}\left(\ket{g_+,g_-}+e^{i(\phi_{L,2}+\phi_{D,2}-\phi_{L,1}-\phi_{D,1})}\ket{g_-,g_+}\right)\ket{1} \right. \nonumber \\ 
    &+ \left.  pe^{i(\phi_{L,1}+\phi_{D,1}+ \phi_{L,2}+\phi_{D,2})}\ket{g_+,g_+}\ket{2}\right].
\end{align}
The fidelity of this state with respect to a two-particle Bell state is bounded by the process that scatters more than one photon, i.e., the maximum achievable fidelity with respect to a Bell state is
\begin{equation}
    F^\text{WPE}_{1,2} = \left|\langle \psi^\phi | \psi_{1,2}\rangle \right|^2\\
    = \frac{2p(1-p)}{2p(1-p)+p^2}.\label{Eq:Fidelity2-1} 
\end{equation}
For example, when using an excitation probability $p=\SI{6}{\percent}$, the maximum attainable fidelity is \SI{96.9}{\percent}. 
The rate of entanglement heralding events is proportional to \begin{align}
    R^\text{WPE}_{1,2} \propto \eta_\text{DET} \left(2p(1-p)+p^2\right), 
\end{align}
where $\eta_\text{DET}$ includes the collection efficiency, fibre transmission, the transmission efficiency of the WPE and the detectors efficiency. It is worth noticing that this rate includes single photon heralding events, but also double photon detection. These are indistinguishable from single photon events and do not create entanglement, reducing the fidelity of the created state. 

This process has been used to demonstrate deterministic distribution of entanglement on NV-centre systems \cite{humphreys2018deterministic},  and for several experiments demonstrating generation of entanglement in trapped-ion systems \cite{slodivcka2013atom,araneda2018interference},
in quantum-dot systems \cite{delteil2016generation,stockill2017phase} in NV-centres \cite{bernien2013heralded,humphreys2018deterministic} and rare-earth on qubits in crystalline structures \cite{ruskuc2024scalable}. It has also been used to create bipartite entanglement in a three-node network \cite{hermans2022qubit}.
In these implementations, the main sources of errors are the finite probability of exciting two atoms at the same time, residual which-path information in the detected photons, distinguishability of the atoms due to recoil, uncertainty in the phase due to residual thermal motion, and the limited fidelity of the coherent operations. Ref.~\cite{hermans2023entangling} shows a detailed analysis of different sources of error for this implementation, and specifically discusses the ones that arise in NV-centre systems.

Notice also that in the  implementation with trapped ions, such as the one presented in \cite{araneda2018interference}, the excitation is done via a non-deterministic Raman process. Since there are other decay channels in the atom's electronic structure, this results in additional errors.

\subsection{Three Nodes}
The photon-number scheme can be generalised to any number of particles. Let us consider the example of three atoms. If three atoms, indexed by $j$, undergo the process described above, the total state of the system is now given by
\begin{align}
           &\phantom\times\left(\sqrt{1-p_1}\ket{g_-,0}_1 + \sqrt{p_1}e^{i\phi_{D,1}+i\phi_{L,1}}\ket{g_+,1}_1\right) \nonumber\\
    &\otimes \left(\sqrt{1-p_2}\ket{g_-,0}_2 + \sqrt{p_2}e^{i\phi_{D,2}+i\phi_{L,2}}\ket{g_+,1}_2\right)\nonumber\\
    &\otimes \left(\sqrt{1-p_3}\ket{g_-,0}_3 + \sqrt{p_3}e^{i\phi_{D,3}+i\phi_{L,3}}\ket{g_+,1}_3\right)
\end{align}
expanding this, we get
\begin{align} 
    &\phantom\times\sqrt{1-p_1}\sqrt{1-p_2}\sqrt{1-p_3}\ket{g_-,g_-,g_,0,0,0}\nonumber\\\nonumber \\
          &+ \sqrt{p_1}\sqrt{1-p_2}\sqrt{1-p_3}e^{i\phi_1}\ket{g_+,g_-,g_-,1,0,0}\nonumber\\
          &+ \sqrt{1-p_1}\sqrt{p_2}\sqrt{1-p_3}e^{i\phi_2}\ket{g_-,g_+,g_-,0,1,0}\nonumber\\
          &+ \sqrt{1-p_1}\sqrt{1-p_2}\sqrt{p_3}e^{i\phi_3}\ket{g_-,g_-,g_+,0,0,1}\nonumber\\\nonumber \\
          &+ \sqrt{p_1}\sqrt{p_2}\sqrt{1-p_3}e^{i\phi_1}e^{i\phi_2}\ket{g_+,g_+,g_-,1,1,0}\nonumber\\
          &+ \sqrt{1-p_1}\sqrt{p_2}\sqrt{p_3}e^{i\phi_2}e^{i\phi_3}\ket{g_-,g_+,g_+,0,1,1}\nonumber\\
          &+ \sqrt{p_1}\sqrt{1-p_2}\sqrt{p_3}e^{i\phi_3}e^{i\phi_1}\ket{g_+,g_-,g_+,1,0,1}\nonumber\\\nonumber \\
           &+\sqrt{p_1}\sqrt{p_2}\sqrt{p_3}e^{i\phi_1}e^{i\phi_2}e^{i\phi_3}\ket{g_+,g_+,g_+,1,1,1},
\end{align}
where $\phi_n = \phi_{L,n}+\phi_{D,n}$. In this case, there are non-vanishing probabilities of emitting 0, 1, 2, or 3 photons. Let us consider that all the atoms are excited with the same probability $p$, so that the state simplifies to
\begin{align}
          &\phantom\pm (1-p)^{3/2}\ket{g_-,g_-,g_,0,0,0} \nonumber\\ \nonumber\\
         & + \sqrt{p}(1-p)e^{i\phi_1}\ket{g_+,g_-,g_-,1,0,0}\nonumber\\
          &+ \sqrt{p}(1-p)e^{i\phi_2}\ket{g_-,g_+,g_-,0,1,0}\nonumber\\
         & + \sqrt{p}(1-p)e^{i\phi_3}\ket{g_-,g_-,g_+,0,0,1}\nonumber\\ \nonumber\\
          &+ p\sqrt{1-p}e^{i\phi_1}e^{i\phi_2}\ket{g_+,g_+,g_-,1,1,0}\nonumber\\
         & + p\sqrt{1-p}e^{i\phi_2}e^{i\phi_3}\ket{g_-,g_+,g_+,0,1,1}\nonumber\\
          &+ p\sqrt{1-p}e^{i\phi_3}e^{i\phi_1}\ket{g_+,g_-,g_+,1,0,1}\nonumber\\ \nonumber\\
          &+ p^{3/2} e^{i\phi_1} e^{i\phi_2}e^{i\phi_3}\ket{g_+,g_+,g_+,1,1,1},
\end{align}

\subsubsection{Single-Excitation W-state}
If we direct the collected photons to a WPE, as shown in Fig.~\ref{fig:cabrillo}b, the emission and detection of a single indistinguishable photon projects, approximately, the state to
\begin{align} 
          \frac{1}{\sqrt{3}}\left(\ket{g_+,g_-,g_-} + e^{i\phi_2}\ket{g_-,g_+,g_-}\right. 
          + \left.e^{i\phi_3}\ket{g_-,g_-,g_+}\right)\label{eq:cab_1_ex},
\end{align}
This multipartite entangled state is called a single excitation W-state. Notice that the optical setup shown in Fig.~\ref{fig:cabrillo}b is a symmetric 3-to-3 multiport, i.e, a tritter \cite{zukowski1997realizable}, allowing for the erasure of the which-path information of the detected photon.  
If we use non-number-resolved photodetectors, the fidelity is limited by the non-negligible probability of not differentiating between one or more photons being detected in a single detector. Thereafter, the state of Eq.~\eqref{eq:cab_1_ex} can be generated with a maximum fidelity of
\begin{align}
    F^\text{WPE}_{1,3} =\frac{3p(1-p)^2}{3p(1-p)^2+3p^2(1-p)+p^3}.
\end{align}
If we consider all the photon emissions that are detected as single clicks in the non-number-resolved photodetectors, the probability of detecting a heralding photon, and thereafter, the rate of entanglement generation is proportional to
\begin{align}
    P^\text{WPE}_{1,3} \propto \eta_{\text{DET}}\left(3p(1-p)^2+3p^2(1-p)+p^3\right).
\end{align}
%hence, if the atom preparation, excitation, and detection time add up to $t_p$, the time required to generate the state of Eq. \eqref{eq:cab_1_ex} is 
%\begin{align}
%    T_E = \frac{t_p}{\eta\left(3p(1-p)^2+3p^2(1-p)+p^3\right)}.
%\end{align}
A demonstration of this scheme for 3-atom entanglement, with two of them in the same node, has been shown in ref.~\cite{ruskuc2024scalable}.

\subsubsection{Two-Excitation W-state}
It is also possible to create an entangled state with a superposition of two atoms being excited, i.e., approximately 
\begin{align}
    & \frac{1}{\sqrt{3}}\left( e^{i\phi_1}e^{i\phi_2}\ket{g_+,g_+,g_-}\right.\nonumber\\
          &+ \left.e^{i\phi_2}e^{i\phi_3}\ket{g_-,g_+,g_+} +  e^{i\phi_1}e^{i\phi_3}\ket{g_+,g_-,g_+} \right).\label{eq:cab_2_ex}   
\end{align}
This process is heralded by the detection of two photons at the output of the WPE. The maximum fidelity with respect to the state of Eq.~\eqref{eq:cab_2_ex} is limited by the non-negligible probability of emitting three photons and detecting just two, i.e., 
\begin{align}
    F^\text{WPE}_{2,3} = \eta_\text{DET}^2\frac{3p^2(1 - p)}{3p^2(1 - p) + p^3}.
\end{align}
The probability of generating this state is proportional to the probability of detecting two photons at the outputs of the WPE. When using the tritter setup of Fig.~\ref{fig:cabrillo}b and using non-number-resolved photo-detectors, the probability of detecting two photons, and therefore the entanglement generation rate is given by
\begin{align}
    P^\text{WPE}_{2,3} = p_{\text{WPE},2,3}\cdot \eta_\text{DET}^2(3p^2(1 - p) + p^3),
\end{align}
where, for the case of the tritter,  $p_{\text{wpe},2,3}=1/3$ is the probability of two photons exiting in two different outputs. The other $2/3$ of the photons bunch together in any of the three detectors with equal probability $2/9$, due to the Hong–Ou–Mandel effect. Thereafter, using number-resolved photodetectors can increase $p_{\text{WPE},2,3}$ to reach unity.

If we choose $p = 6\%$, then the probability of emitting one photon is $\approx$ 16\%, two photons is $\approx$ 0.1\%, and three photons is $2\cdot 10^{-4}$. With 83\% probability, no photons are produced. The maximum fidelity of the one-excitation state using this value is $F^\text{WPE}_{1,3}$ = 93.88\%, while for the 2-excitation entangled state is $F^\text{WPE}_{2,3}$ = 97.91\%. The time required to create the one-excitation state scales linearly with $\eta_\text{DET}$, while the two excitation scales with $\eta_\text{DET}^2$, so there is a strong trade-off between fidelity and entanglement rate to be considered for practical implementations.

\subsection{$N$ Nodes}
The photon-number scheme can be readily extended for any number of particles. If we consider $N$ atoms, all prepared in the same initial superposition, the state after excitation can be written as \cite{wiegner2011quantum}
\begin{align}
    \ket{\Psi} = \sum_{n=0}^{N} \binom{N}{n}^{1/2} p^{n/2} (1-p)^{(N-n)/2}\ket{W_{n,N-n}}\ket{n},
\end{align}
where $\ket{n}$ is the number of photons emitted collectively by the atoms, and $\ket{W_{n,N}}$ are the so-called generalised $W$ states, or generalised Dicke states. They can be defined as
\begin{align}
    \ket{W_{n,N}} = \binom{N}{n}^{-1/2}\sum_k \mathcal{P}_k \left[\ket{S_{n,N-n}}\right], 
\end{align}
where $\mathcal{P}_k$ is the operator that produces all the possible different permutations with an equal number $n$ of particles being in the $\ket{g_+}$ state, and with $\ket{S_{n,N-n}}$ defined by
\begin{align}
    \ket{S_{n,N-n}} = \prod_{\alpha=1}^n e^{i\phi_\alpha}\ket{g_+}_\alpha \prod_{\beta=n+1}^N \ket{g_-}_\beta.
\end{align}
Thereafter, the probability of emitting only one photon is the prefactor of the $\ket{W_{1,N-1}}$ state, and is given by 
\begin{align}
    p_{1,N} = N p (1-p)^{N-1}.
\end{align}
The probability of emitting $n$ photons is given by
\begin{align}
    p_{n,N} = \binom{N}{n} p^n (1-p)^{N-n}.
\end{align}

As before, if we cannot distinguish the number of photons detected, the fidelity is limited by the multi-photon process. If we aim to generate the state $\ket{W_{1,N}}$, the fidelity is limited to
\begin{align}
    F^\text{WPE}_{1,N} = \frac{Np(1-p)^{N-1}}{\sum\limits_{n=1}^N\binom{N}{n} p^n (1-p)^{N-n} }.
\end{align}
The probability of detecting a photon heralding this process is proportional to
\begin{align}
R^\text{WPE}_{1,N} \propto \eta_\text{DET}\sum\limits_{n=1}^N\binom{N}{n} p^n (1-p)^{N-n}.
\end{align}

If we aim to create a state with two excitations, i.e., $\ket{W_{2,N}}$ by detecting two photons, the fidelity is limited by the multi-photon process, which is not distinguishable from the two-photon process due to optical losses, i.e.,
\begin{align}
    F^\text{WPE}_{2,N} = \frac{Np^2(1-p)^{N-2}}{\sum\limits_{n=2}^N\binom{N}{n} p^n (1-p)^{N-n} }.
\end{align}
The probability of detecting two photons heralding this process is proportional to
\begin{align}
R^\text{WPE}_{2,N} \propto \eta^2\sum\limits_{n=2}^N\binom{N}{n} p^n (1-p)^{N-n}.
\end{align}

In general, if we aim to create the $m$ excitations state, i.e., $\ket{W_{m,N-m}}$, the multi-photon processes limit the maximum attainable fidelity to
\begin{align}
    F^\text{WPE}_{m,N} = \frac{Np^m(1-p)^{N-m}}{\sum\limits_{n=m}^N\binom{N}{n} p^n (1-p)^{N-n} }\label{eq:fid_N}.
\end{align}
The probability of detecting $m$ photons heralding this process, and therefore the generation rate of the state $\ket{W}_{m,N}$, is proportional to
\begin{align}
R^\text{WPE}_{m,N} \propto \eta_\text{DET}^m\sum\limits_{n=m}^N\binom{N}{n} p^m (1-p)^{N-n}.\label{eq:rate_N}
\end{align}

\begin{figure}
\centerline{\includegraphics[width=0.8\columnwidth]{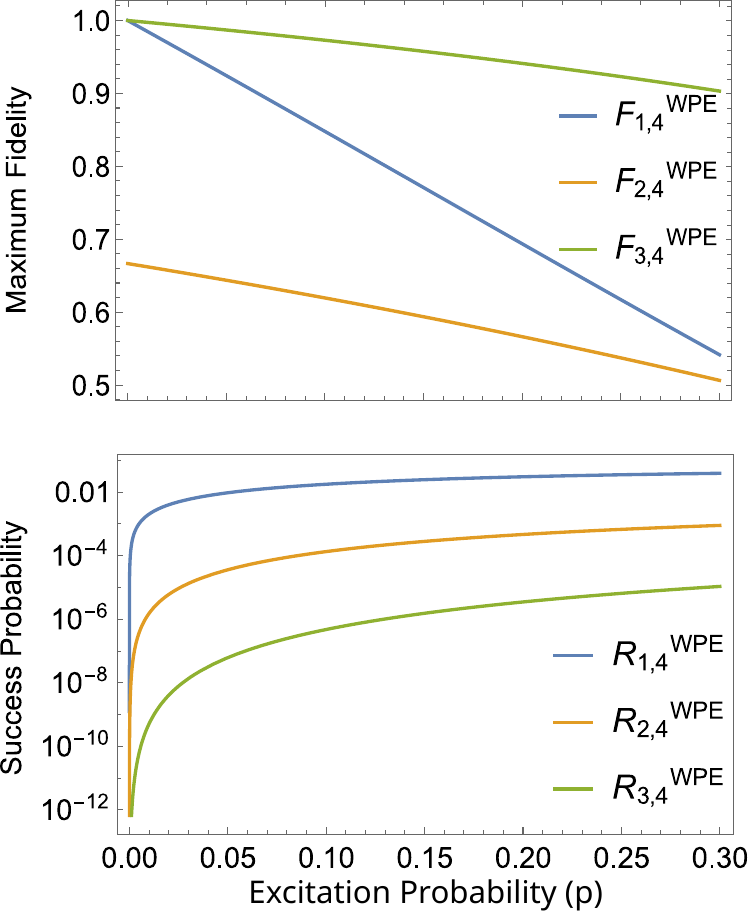}}
\caption{Maximum fidelity and entanglement success rate, for $W$ states with 1, 2 and 3 excitations in a 4-node networks, using the WPE scheme as a function of the excitation probability.  $\eta_\text{DET}$ = 0.05 is used for the plots. }
\label{fig:cabrillo_plots}
\end{figure}

The probability of detecting $n$ photons will depend on the availability of number-resolved photodetectors, and the particular implementation of the $n$-to-$n$ multiport, as in general, we will observe some outputs to be suppressed by the Hong-Ou-Mandel effect. As before, depending on the dimension, symmetric multiports can be built using combinations of beam splitters and tritters (see section \ref{sec:swap_N}). Other alternatives include the use of integrated multi-mode interferometers \cite{peruzzo2011multimode}.
Importantly, the entanglement rate using this scheme scales as $\eta_\text{DET}^m$, and since $m<N$, it is always better than the swapping scheme, which scales as $\eta_\text{DET}^N$. Fig.~\ref{fig:cabrillo_plots} shows the maximum achievable fidelities and entanglement success rates in a 4-node network, for generalised states with 1, 2 and 3 excitations, using $\eta_\text{DET} = 0.05$.

\section{Optical Phase Control}
The fidelity of entangled states is limited by the repeatability and stability of the optical phases of photons traveling through fibres and interferometers. Variations in these optical phases are imprinted on the generated entangled states, causing them to fluctuate over time. While the entanglement swapping scheme and the itinerant photon scheme are resilient to large phase drifts \cite{knollmann2024integrated,knaut2024entanglement}, approaches such as the which-path erasing, single-photon exchange between multiple nodes, and entanglement mapping require sub-wavelength path length stabilisation.

Various methods can actively stabilise these phases. In \cite{slodivcka2012interferometric}, the fluorescence light emitted by atoms during cooling is used to measure phase drift, and a movable mirror compensates for this drift using a sample-and-hold phase lock. In \cite{humphreys2018deterministic}, phase stabilisation is achieved by illuminating the sample substrate and using the reflected light to measure phase drift, with a fibre stretcher employed for compensation. A similar approach is described in \cite{ruskuc2024scalable}.

Even if the phase is unknown, it can be distilled out, provided that phase drifts are slower than the time required for multiple repetitions \cite{delteil2016generation}. These conditions are typically met for short fibre lengths, as phase drift generally occurs on the scale of seconds, while the time required to generate two consecutive entangling events is less than a millisecond under realistic parameters.

Another source of phase uncertainty, particularly in atomic systems, is the intrinsic motion of atoms in trapping potentials. Uncertainty in the atom's position at the moment of absorbing or emitting a photon, causes path-length uncertainties. This uncertainty can be reduced by cooling the atoms to the ground state \cite{slodivcka2013atom}. Additionally, residual motion-qubit entanglement due to recoil after the emission or absorption of single photons can be mitigated by selecting proper absorption and emission directions and, in some cases, by synchronising ion oscillation periods and absorption and emission times \cite{saha2024high}.

\section{Discussion}
We have presented and discussed different schemes for the generation of genuine multipartite remote entanglement over an \(N\)-node quantum network and considered its implementation on different experimental platforms used for quantum networking with matter qubits, including NV-centres, trapped ions, neutral atoms and superconducting qubits. The main advantage of using these schemes in comparison to schemes based solely on the use of bipartite entanglement (see, e.g., \cite{komar2014quantum, mooney2021generation}) is that the single-shot schemes presented here do not require extensive use of local measurement, feed-forward and local entangling gates. Therefore, the desired entangled states can be generated without excessive network latency, and without the need of ancillary qubits or long coherence times.

Typically, in a realistic network node, one could have many qubits, but a limited number of atom-photon interfaces per node. The qubit interacting with the interface will therefore need to be swapped, either physically or logically, during the execution of the scheme. Thereafter, using bipartite-entanglement-based schemes, bipartite entanglement will need to be created in a sequential way between different nodes. For this to work with high fidelity, the coherence time of the individual qubits needs to be much longer than the total time required to establish the desired multipartite state. Although this is feasible for some experimental platforms (see e.g., \cite{drmota2023robust,stas2022robust}), it is still a challenge for most experimental realisations of quantum networks. Furthermore, the overhead on the number of ancillary qubits and the number of local-entangling gates needed when using the bipartite-entanglement-based schemes grows with the number of nodes to entangle, and hence, the fidelity of the involved operations.

All the schemes presented here do not require any extra qubits to create the multi-node entangled state nor local entangling gates, and importantly, do not rely on long coherence times to generate multi-node entanglement because the entanglement is performed in one single step after the initial state preparation of the qubits. For all the heralded schemes presented here, if a try does not generate entanglement, the initial state is prepared again, and a new try is executed. A single try has been demonstrated to be as fast as 1 µs for atomic-based systems \cite{stephenson2020high}, while the coherence time can be, even without the use of a dedicated memory ion, several orders of magnitude longer \cite{drmota2023robust}.

Regarding efficiency, the best scaling with the number of nodes is expected to be achieved using the which-path erasing scheme. However, this scheme also shows the strongest trade-off between multipartite entanglement rate and achievable fidelity (Eqs. \eqref{eq:fid_N} and \eqref{eq:rate_N}), which may render it impractical for the most stringent applications. All other schemes discussed here show entangling rates that depend on \(\eta^N\), where \(N\) is the number of nodes and \(\eta\) is the efficiency of the process. Generally, this will result in slow multi-node entangling rates compared to schemes based solely on bipartite entanglement \cite{mooney2021generation}. However, this limitation diminishes as \(\eta\) approaches 1.

\begin{figure}
\centerline{\includegraphics[width=1.0\columnwidth]{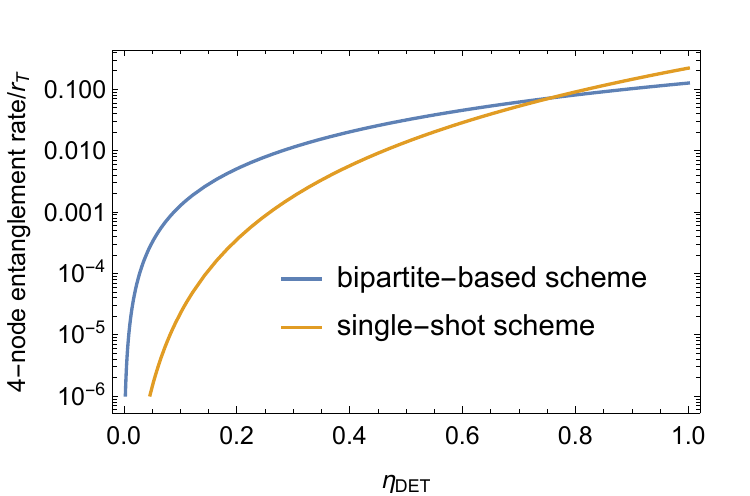}}
\caption{Comparison of multipartite entanglement generation rates in a 4-node network, using bipartite entanglement as a resource, versus the swapping scheme presented in Section \ref{sec:swap}. The horizontal axis shows \(\eta_{\text{DET}}\), the collection and detection efficiency of the emitted single photons.}
\label{fig:comparision}
\end{figure}

Consider, for example, a 4-node network of trapped ions using the parameters presented in \cite{stephenson2020high} for Bell pair generation. In this case, the relevant efficiency \(\eta\) is \(\eta_{\text{DET}}\), the probability of collecting and detecting a single photon emitted by the trapped ions. The generation rate of bipartite entanglement is given by (Eq.~\eqref{Eq:rate_bell_swapp})
\begin{align}
    R_{\text{Bell}} = \frac{1}{2} \eta_{\text{DET}}^2 r_{\text{T}},
\end{align}
where \(r_{\text{T}}\) is the entanglement trial rate. To generate entanglement between four nodes, assuming that each node has at most one ion-photon interface (the Bell state generation cannot be parallelised), and neglecting the time required to perform the Bell measurements and local entangling gates necessary \cite{mooney2021generation}, the rate of 4-node entanglement generation is \(\leq (1/4) R_{\text{Bell}}\), assuming that four rounds of bipartite entanglement is required to generate genuine entanglement between the four nodes (see e.g., \cite{avis2023analysis}).

On the other hand, using the scheme for 4-node entanglement proposed in Appendix \ref{app:Swapp_4_nodes}, the entanglement generation rate is given by
\begin{align}
    R_{\text{Quad}} = \frac{7}{32} \eta_{\text{DET}}^4 r_{\text{T}},
\end{align}
assuming only non-number resolved detectors are available. Figure~\ref{fig:comparision} shows that for \(\eta_{\text{DET}} \gtrapprox 0.76\), the 4-photon scheme is more efficient than the bipartite-based scheme. This \(\eta_{\text{DET}} \gtrapprox 0.76\) is very close to efficiencies already achieved in trapped ion systems \cite{schupp2021interface}. Conditions for the single-shot scheme being advantageous compared to bipartite entanglement-only schemes can be found for all the discussed schemes by varying the corresponding \(\eta\) parameters.

The diverse range of schemes presented here offers promising avenues for achieving multipartite remote entanglement in quantum networks, extending beyond the limitations of using only bipartite remote entanglement. Each scheme carries its unique advantages and challenges, tailored for specific experimental platforms. Further experiments and simulations are needed to understand the impact of various experimental errors on the quality of generated multipartite entangled states. Refs. such as \cite{mooney2021generation, vivoli2019high, tanji2024rate, pfister2016quantum, lougovski2009verifying, hermans2023entangling, vogell2017deterministic} provide valuable insights into relevant calculations and simulations, paving the way for comprehensive exploration and realisation of these schemes in diverse quantum computing, metrology, and sensing architectures.

\section{Acknowledgements}
We thank David Lucas for valuable comments on the manuscript and Tommaso Faorlin for useful discussion. DM acknowledges support from the U.S. Army Research Office (ref.\ W911NF-18-1-0340).
DPN acknowledges support from Merton College, Oxford.
EMA acknowledges support from the U.K. EPSRC ``Quantum Communications'' Hub EP/T001011/1.
RS acknowledges funding from an EPSRC Fellowship EP/W028026/1 and Balliol College, Oxford.
GAM acknowledges support from Wolfson College, Oxford.
This work was supported by the U.K. EPSRC ``Quantum Computing and Simulation'' Hub EP/T001062/1.

\bibliography{bibliography}

\onecolumngrid
\appendix
\section{Entanglement 
swapping for four nodes}
\label{app:Swapp_4_nodes}
Using the entanglement swapping technique presented in section \ref{sec:swap_N}, let us now consider the case of four identical atom-photon entangled pairs, i.e., 
\begin{align}
\label{ES_identical_four_Bell}
&\bigotimes\limits_{n=1}^4 \frac{1}{\sqrt{2}}\left( \ket{0}_{A_n}\ket{0}_{P_n}\pm\ket{1}_{A_n}\ket{1}_{P_n}\right)=\\
&\frac{1}{\sqrt{16}}  \big(  \left|G_0^{+}\right\rangle_{A_1A_2A_3A_4} \left|G_0^{+}\right\rangle_{P_1P_2P_3P_4}\nonumber\\
&+\left|G_0^{-}\right\rangle_{A_1A_2A_3A_4} \left|G_0^{-}\right\rangle_{P_1P_2P_3P_4}\nonumber\\ 
&+ \left|G_1^{+}\right\rangle_{A_1A_2A_3A_4} \left|G_1^{+}\right\rangle_{P_1P_2P_3P_4}\nonumber \\
&\pm\left|G_1^{-}\right\rangle_{A_1A_2A_3A_4} \left|G_1^{-}\right\rangle_{P_1P_2P_3P_4}\nonumber \\ 
&\pm \left|G_2^{+}\right\rangle_{A_1A_2A_3A_4} \left|G_2^{+}\right\rangle_{P_1P_2P_3P_4}\nonumber\\ 
&\pm\left|G_2^{-}\right\rangle_{A_1A_2A_3A_4} \left|G_2^{-}\right\rangle_{P_1P_2P_3P_4}\nonumber\\ 
&+\left|G_3^{+}\right\rangle_{A_1A_2A_3A_4} \left|G_3^{+}\right\rangle_{P_1P_2P_3P_4}\nonumber \\
&+\left|G_3^{-}\right\rangle_{A_1A_2A_3A_4}
\left|G_3^{-}\right\rangle_{P_1P_2P_3P_4}\nonumber \\
&\pm\left|G_4^{+}\right\rangle_{A_1A_2A_3A_4} \left|G_4^{+}\right\rangle_{P_1P_2P_3P_4}\nonumber\\
&\pm\left|G_4^{-}\right\rangle_{A_1A_2A_3A_4} \left|G_4^{-}\right\rangle_{P_1P_2P_3P_4}\nonumber \\
&+\left|G_5^{+}\right\rangle_{A_1A_2A_3A_4} \left|G_5^{+}\right\rangle _{P_1P_2P_3P_4}\nonumber\\
&+\left|G_5^{-}\right\rangle_{A_1A_2A_3A_4} \left|G_5^{-}\right\rangle _{P_1P_2P_3P_4}\nonumber\\ 
&+\left|G_6^{+}\right\rangle_{A_1A_2A_3A_4} \left|G_6^{+}\right\rangle _{P_1P_2P_3P_4}\nonumber\\
&+\left|G_6^{-}\right\rangle_{A_1A_2A_3A_4} \left|G_6^{-}\right\rangle_{P_1P_2P_3P_4}\nonumber \\
&\pm\left|G_7^{+}\right\rangle_{A_1A_2A_3A_4} \left|G_7^{+}\right\rangle_{P_1P_2P_3P_4}\nonumber \\
&\pm \left|G_7^{-}\right\rangle_{A_1A_2A_3A_4} \left|G_7^{-}\right\rangle_{P_1P_2P_3P_4}\big), \label{eq:4poly}
\end{align}
with
\begin{align}
   \ket{G_0^{\pm}} = \frac{1}{\sqrt{2}}\big(\ket{0000} \pm \ket{1111} \big),\\
    \ket{G_1^{\pm}} = \frac{1}{\sqrt{2}}\big(\ket{0001} \pm \ket{1110} \big),\\
    \ket{G_2^{\pm}} = \frac{1}{\sqrt{2}}\big(\ket{0010} \pm \ket{1101} \big),\\
     \ket{G_3^{\pm}} = \frac{1}{\sqrt{2}}\big(\ket{0011} \pm \ket{1100} \big),\\
      \ket{G_4^{\pm}} = \frac{1}{\sqrt{2}}\big(\ket{0100} \pm \ket{1011} \big),\\
    \ket{G_5^{\pm}} = \frac{1}{\sqrt{2}}\big(\ket{0101} \pm \ket{1010} \big)\\
    \ket{G_6^{\pm}} = \frac{1}{\sqrt{2}}\big(\ket{0110} \pm \ket{1001} \big),\\
     \ket{G_7^{\pm}} = \frac{1}{\sqrt{2}}\big(\ket{0111} \pm \ket{1000} \big), \label{eq:Gs4}
\end{align}
where the $\{\ket{G_n^\pm}\}$ states correspond now to 4-particle entangled states of the GHZ class.
As before, by using a 4-photon generalized BSA it is possible to generate 4-particle entanglement. 

An example of such a 4-photon generalized BSA using photon-polarisation encoding, is shown in Fig.~\ref{fig:GBSA_2_3_4}c. This BSA is a 4-to-4 symmetric multiport \cite{zukowski1997realizable,lim2005multiphoton}: a single photon entering any of the input ports has the same probability of exiting at any of the output ports, erasing the information about which atom emitted that photon. Notice that this so-called quarter uses 50:50 beams splitters only \cite{zukowski1997realizable}. The creation of entanglement of 4-particle entanglement is heralded by four coincidence clicks in some particular photon detection patterns. A detailed analysis for the different detection cases is shown in appendix \ref{app:quarters}.
If a particular implementation of the 4-photon BSA projects the atomic states into an entangled state with probability $p_{\text{BSA}_4}$, the probability of generating a 4-particle entangled state is then
\begin{align}
    R^{\text{SW}}_4 \propto p_{\text{BSA}_4}\eta_\text{DET}^4.
    \label{Eq:rate_bell_swapp}
\end{align}
%and the average required time to generate such a state is
%\begin{align}
%    T_E = \frac{t_p}{p_{BSA_4}\eta^4}
%\end{align}
The implementation of the 4-photon generalised BSA shown in Fig.~\ref{fig:GBSA_2_3_4}c, a called quarter, can generate 4-particle remote entanglement with probability $p_{\text{BSA}_4} = 7/32$ upon coincident detection of single photons at four different detectors. When using number-resolved detectors, and heralding in any combination of four photons detected, the probability of generating entanglement is $7/8$, see appendix \ref{app:quarters} for details. Note that, depending on the heralding detection patterns, not all the generated 4-particle entangled states are one of the $\{\ket{G_n^\pm}\} $ states, but more general 4-particle entangled states of the GHZ and W classes. Importantly, if we attempt to produce entanglement between only two nodes using the quarter shown in Fig.~\ref{fig:GBSA_2_3_4}c, by sending two  photons and detecting two photons at the output, a Bell state 
 between the two nodes is produced also with the maximum probability allowed, i.e., 1/2$\eta_\text{DET}^2$. This enables the production of entanglement of any two in a 4-node without the need of an additional optical switch. The same applies for entangling 3 nodes. 

Using the alternative generalised BSA presented in ref.~\cite{vivoli2019high}, a 4-particle GHZ state can be generated with probability proportional to $p_{\text{BSA}_4} = 1/16$. Other proposed interferometer structures can be found in  ref.~\cite{tanji2024rate}

\section{Entanglement swapping using a quarter}
\label{app:quarters}
\begin{figure}[b!]
\centerline{\includegraphics[width=0.3\columnwidth]{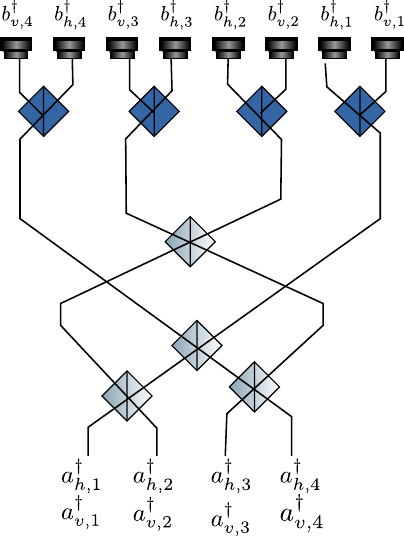}}
\caption{Input and output for a specific realisation of a Quarter.}
\label{fig:quarter}
\end{figure}
We consider now the case of entanglement swapping for a 4-node network using a quarter (see Fig.~\ref{fig:quarter}). The quarter is unitary transformation from the input modes $a$ to the output modes $b$, such that the creation operators for each mode are related through
\begin{align}
\left(
\begin{array}{c}
 b^\dagger_{\text{\{h,v\}1}} \\
 b^\dagger_{\text{\{h,v\}2}} \\
 b^\dagger_{\text{\{h,v\}3}} \\
b^\dagger_{\text{\{h,v\}4}} \\
\end{array}
\right)= Q \cdot \left(
\begin{array}{c}
 a^\dagger_{\text{\{h,v\}1}} \\
 a^\dagger_{\text{\{h,v\}2}} \\
 a^\dagger_{\text{\{h,v\}3}} \\
 a^\dagger_{\text{\{h,v\}4}} \\
\end{array}
\right).
\end{align}
For example,  $a^\dagger_{\text{\{h\}1}}$ is the creation operator for a $h$-polarized photon in the input one.
The unitary matrix representing the transformation exerted by the interferometer of Fig.\ref{fig:GBSA_2_3_4}c is the composition of the unitary transformations of a beam splitters acting on different sequentially on different modes
\begin{align}
Q =    \left(
\begin{array}{cccc}
 1 & 0 & 0 & 0 \\
 0 & \frac{1}{\sqrt{2}} & \frac{i}{\sqrt{2}} & 0 \\
 0 & \frac{i}{\sqrt{2}} & \frac{1}{\sqrt{2}} & 0 \\
 0 & 0 & 0 & 1 \\
\end{array}
\right) \cdot
\left(
\begin{array}{cccc}
 \frac{1}{\sqrt{2}} & 0 & 0 & \frac{i}{\sqrt{2}} \\
 0 & 1 & 0 & 0 \\
 0 & 0 & 1 & 0 \\
 \frac{i}{\sqrt{2}} & 0 & 0 & \frac{1}{\sqrt{2}} \\
\end{array}
\right) \cdot
\left(
\begin{array}{cccc}
 1 & 0 & 0 & 0 \\
 0 & 1 & 0 & 0 \\
 0 & 0 & \frac{1}{\sqrt{2}} & \frac{i}{\sqrt{2}} \\
 0 & 0 & \frac{i}{\sqrt{2}} & \frac{1}{\sqrt{2}} \\
\end{array}
\right)\cdot
\left(
\begin{array}{cccc}
 \frac{1}{\sqrt{2}} & \frac{i}{\sqrt{2}} & 0 & 0 \\
 \frac{i}{\sqrt{2}} & \frac{1}{\sqrt{2}} & 0 & 0 \\
 0 & 0 & 1 & 0 \\
 0 & 0 & 0 & 1 \\
\end{array}
\right).
\end{align}
Notice that this is a particular realisation of a quarter, and in principle, arbitrary phase shift could be placed in any path of the interferometer, changing this matrix and the obtained state that we will listen below. The phases could be optimised to achieve certain state with high probability.
We condition on the detection of some combination of clicks at the output states $b$, so we need to write the input states in terms of output states. This is done through the inverse matrix
\begin{align}
\left(
\begin{array}{c}
 a^\dagger_{\text{\{h,v\}1}} \\
 a^\dagger_{\text{\{h,v\}2}} \\
 a^\dagger_{\text{\{h,v\}3}} \\
a^\dagger_{\text{\{h,v\}4}} \\
\end{array}
\right)= Q^{-1} \cdot \left(
\begin{array}{c}
 b^\dagger_{\text{\{h,v\}1}} \\
 b^\dagger_{\text{\{h,v\}2}} \\
 b^\dagger_{\text{\{h,v\}3}} \\
 b^\dagger_{\text{\{h,v\}4}} \\
\end{array}
\right)\label{eq:quarter_relation}
\end{align}
with
\begin{align}
    Q^{-1} = \left(
\begin{array}{cccc}
 \frac{1}{2} & -\frac{i}{2} & -\frac{1}{2} & -\frac{i}{2} \\
 -\frac{i}{2} & \frac{1}{2} & -\frac{i}{2} & -\frac{1}{2} \\
 -\frac{1}{2} & -\frac{i}{2} & \frac{1}{2} & -\frac{i}{2} \\
 -\frac{i}{2} & -\frac{1}{2} & -\frac{i}{2} & \frac{1}{2} \\
\end{array}
\right).
\end{align}

Any of the input states of the interferometer (Eq. \ref{eq:Gs4}) can be written using the specific combination of creation operators, for example
\begin{align}
    \frac{1}{\sqrt{2}}\left(\ket{0101}_{P_1,P_2,P_3,P4}+\ket{1010}_{P_1,P_2,P_3,P4}\right) &= \frac{1}{\sqrt{4}} \left(\ket{hvhv}+\ket{vhvh}\right)\nonumber\\
    &= \frac{1}{\sqrt{2}}(a^\dagger_{h,1}a^\dagger_{v,2}a^\dagger_{h,3}a^\dagger_{v,4}+a^\dagger_{v,1}a^\dagger_{h,2}a^\dagger_{v,3}a^\dagger_{h,4})\ket{0}.
\end{align}
Then, each of the input field operators $a^\dagger$ operators can be written using Eq. \ref{eq:quarter_relation} in terms of the output operators of the GBSA. Then we can write the full extension of Eq. \ref{eq:4poly}, and see how different output patterns, i.e., photon detection of four photons patterns in the eight detectors, correspond to the different projection of the atomic states. If we consider that the initial state of the atom-photon pairs is given by
\begin{align}
    \bigotimes\limits_{n=1}^4 \frac{1}{\sqrt{2}}\left( \ket{0}_{A_n}\ket{0}_{P_n} + \ket{1}_{A_n}\ket{1}_{P_n} \right),
\end{align}
And we consider detection patterns of four photons, with at most one click per detector, the possible patterns are:

\begin{eqnarray*}
%\arraycolsep=1.4pt\def\arraystretch{2.2}
%\begin{array}{ccc}
\text{\underline{Detection Pattern}} & \text{\underline{State}} & \text{\underline{Probability}} \\
 b_{\text{h1}} b_{\text{h2}} b_{\text{h3}} b_{\text{h4}} & |0000\rangle  &
   \frac{1}{64} \\
 b_{\text{h1}} b_{\text{h2}} b_{\text{h3}} b_{\text{v1}} & -\frac{1}{2} i
   (|0001\rangle +|0010\rangle -|0100\rangle -|1000\rangle ) & \frac{1}{256} \\
 b_{\text{h1}} b_{\text{h2}} b_{\text{h3}} b_{\text{v2}} & \frac{1}{2}
   (-|0001\rangle +|0010\rangle -|0100\rangle +|1000\rangle ) & \frac{1}{256}
   \\
 b_{\text{h1}} b_{\text{h2}} b_{\text{h3}} b_{\text{v3}} & -\frac{1}{2} i
   (|0001\rangle -|0010\rangle -|0100\rangle +|1000\rangle ) & \frac{1}{256} \\
 b_{\text{h1}} b_{\text{h2}} b_{\text{h3}} b_{\text{v4}} & \frac{1}{2}
   (|0001\rangle +|0010\rangle +|0100\rangle +|1000\rangle ) & \frac{1}{256} \\
 b_{\text{h1}} b_{\text{h2}} b_{\text{h4}} b_{\text{v1}} & \frac{1}{2}
   (|0001\rangle -|0010\rangle +|0100\rangle -|1000\rangle ) & \frac{1}{256} \\
 b_{\text{h1}} b_{\text{h2}} b_{\text{h4}} b_{\text{v2}} & -\frac{1}{2} i
   (|0001\rangle +|0010\rangle -|0100\rangle -|1000\rangle ) & \frac{1}{256} \\
 b_{\text{h1}} b_{\text{h2}} b_{\text{h4}} b_{\text{v3}} & \frac{1}{2}
   (|0001\rangle +|0010\rangle +|0100\rangle +|1000\rangle ) & \frac{1}{256} \\
 b_{\text{h1}} b_{\text{h2}} b_{\text{h4}} b_{\text{v4}} & \frac{1}{2} i
   (|0001\rangle -|0010\rangle -|0100\rangle +|1000\rangle ) & \frac{1}{256} \\
 b_{\text{h1}} b_{\text{h2}} b_{\text{v1}} b_{\text{v2}} & \frac{|0110\rangle
   +|1001\rangle }{\sqrt{2}} & \frac{1}{128} \\
 b_{\text{h1}} b_{\text{h2}} b_{\text{v3}} b_{\text{v4}} & \frac{|0110\rangle
   +|1001\rangle }{\sqrt{2}} & \frac{1}{128} \\
 b_{\text{h1}} b_{\text{h3}} b_{\text{h4}} b_{\text{v1}} & \frac{1}{2} i
   (|0001\rangle -|0010\rangle -|0100\rangle +|1000\rangle ) & \frac{1}{256} \\
 b_{\text{h1}} b_{\text{h3}} b_{\text{h4}} b_{\text{v2}} & \frac{1}{2}
   (|0001\rangle +|0010\rangle +|0100\rangle +|1000\rangle ) & \frac{1}{256} \\
 b_{\text{h1}} b_{\text{h3}} b_{\text{h4}} b_{\text{v3}} & \frac{1}{2} i
   (|0001\rangle +|0010\rangle -|0100\rangle -|1000\rangle ) & \frac{1}{256} \\
 b_{\text{h1}} b_{\text{h3}} b_{\text{h4}} b_{\text{v4}} & \frac{1}{2}
   (-|0001\rangle +|0010\rangle -|0100\rangle +|1000\rangle ) & \frac{1}{256}
   \\ 
 b_{\text{h1}} b_{\text{h3}} b_{\text{v1}} b_{\text{v3}} & \frac{-|0101\rangle
   -|1010\rangle }{\sqrt{2}} & \frac{1}{128} \\
 b_{\text{h1}} b_{\text{h3}} b_{\text{v2}} b_{\text{v4}} & \frac{|0101\rangle
   +|1010\rangle }{\sqrt{2}} & \frac{1}{128} \\
 b_{\text{h1}} b_{\text{h4}} b_{\text{v1}} b_{\text{v4}} & \frac{|0011\rangle
   +|1100\rangle }{\sqrt{2}} & \frac{1}{128} \\
 b_{\text{h1}} b_{\text{h4}} b_{\text{v2}} b_{\text{v3}} & \frac{|0011\rangle
   +|1100\rangle }{\sqrt{2}} & \frac{1}{128} \\
 b_{\text{h1}} b_{\text{v1}} b_{\text{v2}} b_{\text{v3}} & \frac{1}{2} i
   (|0111\rangle +|1011\rangle -|1101\rangle -|1110\rangle ) & \frac{1}{256} \\
 b_{\text{h1}} b_{\text{v1}} b_{\text{v2}} b_{\text{v4}} & \frac{1}{2}
   (-|0111\rangle +|1011\rangle -|1101\rangle +|1110\rangle ) & \frac{1}{256}
   \\
 b_{\text{h1}} b_{\text{v1}} b_{\text{v3}} b_{\text{v4}} & \frac{1}{2} i
   (|0111\rangle -|1011\rangle -|1101\rangle +|1110\rangle ) & \frac{1}{256} \\
 b_{\text{h1}} b_{\text{v2}} b_{\text{v3}} b_{\text{v4}} & \frac{1}{2}
   (|0111\rangle +|1011\rangle +|1101\rangle +|1110\rangle ) & \frac{1}{256} \\
 b_{\text{h2}} b_{\text{h3}} b_{\text{h4}} b_{\text{v1}} & \frac{1}{2}
   (|0001\rangle +|0010\rangle +|0100\rangle +|1000\rangle ) & \frac{1}{256} \\
 b_{\text{h2}} b_{\text{h3}} b_{\text{h4}} b_{\text{v2}} & -\frac{1}{2} i
   (|0001\rangle -|0010\rangle -|0100\rangle +|1000\rangle ) & \frac{1}{256} \\
 b_{\text{h2}} b_{\text{h3}} b_{\text{h4}} b_{\text{v3}} & \frac{1}{2}
   (|0001\rangle -|0010\rangle +|0100\rangle -|1000\rangle ) & \frac{1}{256} \\
 b_{\text{h2}} b_{\text{h3}} b_{\text{h4}} b_{\text{v4}} & \frac{1}{2} i
   (|0001\rangle +|0010\rangle -|0100\rangle -|1000\rangle ) & \frac{1}{256} \\
 b_{\text{h2}} b_{\text{h3}} b_{\text{v1}} b_{\text{v4}} & \frac{|0011\rangle
   +|1100\rangle }{\sqrt{2}} & \frac{1}{128} \\
 b_{\text{h2}} b_{\text{h3}} b_{\text{v2}} b_{\text{v3}} & \frac{|0011\rangle
   +|1100\rangle }{\sqrt{2}} & \frac{1}{128} \\
 b_{\text{h2}} b_{\text{h4}} b_{\text{v1}} b_{\text{v3}} & \frac{|0101\rangle
   +|1010\rangle }{\sqrt{2}} & \frac{1}{128} \\
 b_{\text{h2}} b_{\text{h4}} b_{\text{v2}} b_{\text{v4}} & \frac{-|0101\rangle
   -|1010\rangle }{\sqrt{2}} & \frac{1}{128} \\
 b_{\text{h2}} b_{\text{v1}} b_{\text{v2}} b_{\text{v3}} & \frac{1}{2}
   (|0111\rangle -|1011\rangle +|1101\rangle -|1110\rangle ) & \frac{1}{256} \\
 b_{\text{h2}} b_{\text{v1}} b_{\text{v2}} b_{\text{v4}} & \frac{1}{2} i
   (|0111\rangle +|1011\rangle -|1101\rangle -|1110\rangle ) & \frac{1}{256} \\
 b_{\text{h2}} b_{\text{v1}} b_{\text{v3}} b_{\text{v4}} & \frac{1}{2}
   (|0111\rangle +|1011\rangle +|1101\rangle +|1110\rangle ) & \frac{1}{256} \\
 b_{\text{h2}} b_{\text{v2}} b_{\text{v3}} b_{\text{v4}} & -\frac{1}{2} i
   (|0111\rangle -|1011\rangle -|1101\rangle +|1110\rangle ) & \frac{1}{256} \\
 b_{\text{h3}} b_{\text{h4}} b_{\text{v1}} b_{\text{v2}} & \frac{|0110\rangle
   +|1001\rangle }{\sqrt{2}} & \frac{1}{128} \\
 b_{\text{h3}} b_{\text{h4}} b_{\text{v3}} b_{\text{v4}} & \frac{|0110\rangle
   +|1001\rangle }{\sqrt{2}} & \frac{1}{128} \\
 b_{\text{h3}} b_{\text{v1}} b_{\text{v2}} b_{\text{v3}} & -\frac{1}{2} i
   (|0111\rangle -|1011\rangle -|1101\rangle +|1110\rangle ) & \frac{1}{256} \\
 b_{\text{h3}} b_{\text{v1}} b_{\text{v2}} b_{\text{v4}} & \frac{1}{2}
   (|0111\rangle +|1011\rangle +|1101\rangle +|1110\rangle ) & \frac{1}{256} \\
 b_{\text{h3}} b_{\text{v1}} b_{\text{v3}} b_{\text{v4}} & \frac{1}{2} i
   (-|0111\rangle -|1011\rangle +|1101\rangle +|1110\rangle ) & \frac{1}{256}
   \\
 b_{\text{h3}} b_{\text{v2}} b_{\text{v3}} b_{\text{v4}} & \frac{1}{2}
   (-|0111\rangle +|1011\rangle -|1101\rangle +|1110\rangle ) & \frac{1}{256}
   \\
 b_{\text{h4}} b_{\text{v1}} b_{\text{v2}} b_{\text{v3}} & \frac{1}{2}
   (|0111\rangle +|1011\rangle +|1101\rangle +|1110\rangle ) & \frac{1}{256} \\
 b_{\text{h4}} b_{\text{v1}} b_{\text{v2}} b_{\text{v4}} & \frac{1}{2} i
   (|0111\rangle -|1011\rangle -|1101\rangle +|1110\rangle ) & \frac{1}{256} \\
 b_{\text{h4}} b_{\text{v1}} b_{\text{v3}} b_{\text{v4}} & \frac{1}{2}
   (|0111\rangle -|1011\rangle +|1101\rangle -|1110\rangle ) & \frac{1}{256} \\
 b_{\text{h4}} b_{\text{v2}} b_{\text{v3}} b_{\text{v4}} & \frac{1}{2} i
   (-|0111\rangle -|1011\rangle +|1101\rangle +|1110\rangle ) & \frac{1}{256}
   \\
 b_{\text{v1}} b_{\text{v2}} b_{\text{v3}} b_{\text{v4}} & |1111\rangle  &
   \frac{1}{64} \\
%\end{array}
\end{eqnarray*}

\newpage
If we consider at most two clicks per detector, the possible patterns are:
\begin{eqnarray*}
\text{\underline{Detection Pattern}} & \text{\underline{State}} & \text{\underline{Probability}} \\
 b_{\text{h1}}^2 b_{\text{h2}}^2 & |0000\rangle  & \frac{1}{256} \\
 b_{\text{h1}}^2 b_{\text{h2}} b_{\text{v1}} & \frac{1}{2} i (|0001\rangle
   -|0010\rangle -|0100\rangle +|1000\rangle ) & \frac{1}{512} \\
 b_{\text{h1}}^2 b_{\text{h2}} b_{\text{v2}} & \frac{1}{2} (|0001\rangle
   +|0010\rangle +|0100\rangle +|1000\rangle ) & \frac{1}{512} \\
 b_{\text{h1}}^2 b_{\text{h2}} b_{\text{v3}} & \frac{1}{2} i (|0001\rangle
   +|0010\rangle -|0100\rangle -|1000\rangle ) & \frac{1}{512} \\
 b_{\text{h1}}^2 b_{\text{h2}} b_{\text{v4}} & \frac{1}{2} (-|0001\rangle
   +|0010\rangle -|0100\rangle +|1000\rangle ) & \frac{1}{512} \\
 b_{\text{h1}}^2 b_{\text{h3}}^2 & -|0000\rangle  & \frac{1}{256} \\
 b_{\text{h1}}^2 b_{\text{h3}} b_{\text{v1}} & \frac{1}{2} (-|0001\rangle
   +|0010\rangle -|0100\rangle +|1000\rangle ) & \frac{1}{512} \\
 b_{\text{h1}}^2 b_{\text{h3}} b_{\text{v2}} & \frac{1}{2} i (|0001\rangle
   +|0010\rangle -|0100\rangle -|1000\rangle ) & \frac{1}{512} \\
 b_{\text{h1}}^2 b_{\text{h3}} b_{\text{v3}} & \frac{1}{2} (-|0001\rangle
   -|0010\rangle -|0100\rangle -|1000\rangle ) & \frac{1}{512} \\
 b_{\text{h1}}^2 b_{\text{h3}} b_{\text{v4}} & -\frac{1}{2} i (|0001\rangle
   -|0010\rangle -|0100\rangle +|1000\rangle ) & \frac{1}{512} \\
 b_{\text{h1}}^2 b_{\text{h4}}^2 & |0000\rangle  & \frac{1}{256} \\
 b_{\text{h1}}^2 b_{\text{h4}} b_{\text{v1}} & -\frac{1}{2} i (|0001\rangle
   +|0010\rangle -|0100\rangle -|1000\rangle ) & \frac{1}{512} \\
 b_{\text{h1}}^2 b_{\text{h4}} b_{\text{v2}} & \frac{1}{2} (-|0001\rangle
   +|0010\rangle -|0100\rangle +|1000\rangle ) & \frac{1}{512} \\
 b_{\text{h1}}^2 b_{\text{h4}} b_{\text{v3}} & -\frac{1}{2} i (|0001\rangle
   -|0010\rangle -|0100\rangle +|1000\rangle ) & \frac{1}{512} \\
 b_{\text{h1}}^2 b_{\text{h4}} b_{\text{v4}} & \frac{1}{2} (|0001\rangle
   +|0010\rangle +|0100\rangle +|1000\rangle ) & \frac{1}{512} \\
 b_{\text{h1}}^2 b_{\text{v1}}^2 & \frac{|0011\rangle +|0101\rangle
   +|0110\rangle +|1001\rangle +|1010\rangle +|1100\rangle }{\sqrt{6}} &
   \frac{3}{512} \\
 b_{\text{h1}}^2 b_{\text{v1}} b_{\text{v2}} & \frac{i (|0110\rangle
   -|1001\rangle )}{\sqrt{2}} & \frac{1}{256} \\
 b_{\text{h1}}^2 b_{\text{v1}} b_{\text{v3}} & \frac{|0101\rangle -|1010\rangle
   }{\sqrt{2}} & \frac{1}{256} \\
 b_{\text{h1}}^2 b_{\text{v1}} b_{\text{v4}} & \frac{i (|0011\rangle
   -|1100\rangle )}{\sqrt{2}} & \frac{1}{256} \\
 b_{\text{h1}}^2 b_{\text{v2}}^2 & \frac{|0011\rangle +|0101\rangle
   -|0110\rangle -|1001\rangle +|1010\rangle +|1100\rangle }{\sqrt{6}} &
   \frac{3}{512} \\
 b_{\text{h1}}^2 b_{\text{v2}} b_{\text{v3}} & \frac{i (|0011\rangle
   -|1100\rangle )}{\sqrt{2}} & \frac{1}{256} \\
 b_{\text{h1}}^2 b_{\text{v2}} b_{\text{v4}} & \frac{|1010\rangle -|0101\rangle
   }{\sqrt{2}} & \frac{1}{256} \\
 b_{\text{h1}}^2 b_{\text{v3}}^2 & \frac{-|0011\rangle +|0101\rangle
   -|0110\rangle -|1001\rangle +|1010\rangle -|1100\rangle }{\sqrt{6}} &
   \frac{3}{512} \\
 b_{\text{h1}}^2 b_{\text{v3}} b_{\text{v4}} & \frac{i (|0110\rangle
   -|1001\rangle )}{\sqrt{2}} & \frac{1}{256} \\
 b_{\text{h1}}^2 b_{\text{v4}}^2 & \frac{-|0011\rangle +|0101\rangle
   +|0110\rangle +|1001\rangle +|1010\rangle -|1100\rangle }{\sqrt{6}} &
   \frac{3}{512} \\
 b_{\text{h1}} b_{\text{h2}}^2 b_{\text{v1}} & \frac{1}{2} (|0001\rangle
   +|0010\rangle +|0100\rangle +|1000\rangle ) & \frac{1}{512} \\
 b_{\text{h1}} b_{\text{h2}}^2 b_{\text{v2}} & -\frac{1}{2} i (|0001\rangle
   -|0010\rangle -|0100\rangle +|1000\rangle ) & \frac{1}{512} \\
 b_{\text{h1}} b_{\text{h2}}^2 b_{\text{v3}} & \frac{1}{2} (|0001\rangle
   -|0010\rangle +|0100\rangle -|1000\rangle ) & \frac{1}{512} \\
 b_{\text{h1}} b_{\text{h2}}^2 b_{\text{v4}} & \frac{1}{2} i (|0001\rangle
   +|0010\rangle -|0100\rangle -|1000\rangle ) & \frac{1}{512} \\
 b_{\text{h1}} b_{\text{h2}} b_{\text{v1}}^2 & -\frac{i (|0110\rangle
   -|1001\rangle )}{\sqrt{2}} & \frac{1}{256} \\
 b_{\text{h1}} b_{\text{h2}} b_{\text{v2}}^2 & \frac{i (|0110\rangle
   -|1001\rangle )}{\sqrt{2}} & \frac{1}{256} \\
 b_{\text{h1}} b_{\text{h2}} b_{\text{v3}}^2 & \frac{i (|0110\rangle
   -|1001\rangle )}{\sqrt{2}} & \frac{1}{256} \\
 b_{\text{h1}} b_{\text{h2}} b_{\text{v4}}^2 & -\frac{i (|0110\rangle
   -|1001\rangle )}{\sqrt{2}} & \frac{1}{256} \\
 b_{\text{h1}} b_{\text{h3}}^2 b_{\text{v1}} & \frac{1}{2} (-|0001\rangle
   -|0010\rangle -|0100\rangle -|1000\rangle ) & \frac{1}{512} \\
 b_{\text{h1}} b_{\text{h3}}^2 b_{\text{v2}} & \frac{1}{2} i (|0001\rangle
   -|0010\rangle -|0100\rangle +|1000\rangle ) & \frac{1}{512} \\
 b_{\text{h1}} b_{\text{h3}}^2 b_{\text{v3}} & \frac{1}{2} (-|0001\rangle
   +|0010\rangle -|0100\rangle +|1000\rangle ) & \frac{1}{512} \\
 b_{\text{h1}} b_{\text{h3}}^2 b_{\text{v4}} & -\frac{1}{2} i (|0001\rangle
   +|0010\rangle -|0100\rangle -|1000\rangle ) & \frac{1}{512} \\
 b_{\text{h1}} b_{\text{h3}} b_{\text{v1}}^2 & \frac{|1010\rangle -|0101\rangle
   }{\sqrt{2}} & \frac{1}{256} \\
 b_{\text{h1}} b_{\text{h3}} b_{\text{v2}}^2 & \frac{|1010\rangle -|0101\rangle
   }{\sqrt{2}} & \frac{1}{256} \\
 b_{\text{h1}} b_{\text{h3}} b_{\text{v3}}^2 & \frac{|1010\rangle -|0101\rangle
   }{\sqrt{2}} & \frac{1}{256} \\
 b_{\text{h1}} b_{\text{h3}} b_{\text{v4}}^2 & \frac{|1010\rangle -|0101\rangle
   }{\sqrt{2}} & \frac{1}{256} \\
 b_{\text{h1}} b_{\text{h4}}^2 b_{\text{v1}} & \frac{1}{2} (|0001\rangle
   +|0010\rangle +|0100\rangle +|1000\rangle ) & \frac{1}{512} \\
 b_{\text{h1}} b_{\text{h4}}^2 b_{\text{v2}} & -\frac{1}{2} i (|0001\rangle
   -|0010\rangle -|0100\rangle +|1000\rangle ) & \frac{1}{512} \\
 b_{\text{h1}} b_{\text{h4}}^2 b_{\text{v3}} & \frac{1}{2} (|0001\rangle
   -|0010\rangle +|0100\rangle -|1000\rangle ) & \frac{1}{512} \\
 b_{\text{h1}} b_{\text{h4}}^2 b_{\text{v4}} & \frac{1}{2} i (|0001\rangle
   +|0010\rangle -|0100\rangle -|1000\rangle ) & \frac{1}{512} \\
 b_{\text{h1}} b_{\text{h4}} b_{\text{v1}}^2 & -\frac{i (|0011\rangle
   -|1100\rangle )}{\sqrt{2}} & \frac{1}{256} \\
 b_{\text{h1}} b_{\text{h4}} b_{\text{v2}}^2 & -\frac{i (|0011\rangle
   -|1100\rangle )}{\sqrt{2}} & \frac{1}{256} \\
 b_{\text{h1}} b_{\text{h4}} b_{\text{v3}}^2 & \frac{i (|0011\rangle
   -|1100\rangle )}{\sqrt{2}} & \frac{1}{256} \\
 b_{\text{h1}} b_{\text{h4}} b_{\text{v4}}^2 & \frac{i (|0011\rangle
   -|1100\rangle )}{\sqrt{2}} & \frac{1}{256} \\
 b_{\text{h1}} b_{\text{v1}}^2 b_{\text{v2}} & \frac{1}{2} i (|0111\rangle
   -|1011\rangle -|1101\rangle +|1110\rangle ) & \frac{1}{512} \\
 b_{\text{h1}} b_{\text{v1}}^2 b_{\text{v3}} & \frac{1}{2} (|0111\rangle
   -|1011\rangle +|1101\rangle -|1110\rangle ) & \frac{1}{512} \\
 b_{\text{h1}} b_{\text{v1}}^2 b_{\text{v4}} & \frac{1}{2} i (|0111\rangle
   +|1011\rangle -|1101\rangle -|1110\rangle ) & \frac{1}{512} \\
 b_{\text{h1}} b_{\text{v1}} b_{\text{v2}}^2 & \frac{1}{2} (|0111\rangle
   +|1011\rangle +|1101\rangle +|1110\rangle ) & \frac{1}{512} \\
 b_{\text{h1}} b_{\text{v1}} b_{\text{v3}}^2 & \frac{1}{2} (-|0111\rangle
   -|1011\rangle -|1101\rangle -|1110\rangle ) & \frac{1}{512} \\
 b_{\text{h1}} b_{\text{v1}} b_{\text{v4}}^2 & \frac{1}{2} (|0111\rangle
   +|1011\rangle +|1101\rangle +|1110\rangle ) & \frac{1}{512} \\
 b_{\text{h1}} b_{\text{v2}}^2 b_{\text{v3}} & \frac{1}{2} (-|0111\rangle
   +|1011\rangle -|1101\rangle +|1110\rangle ) & \frac{1}{512} \\
 b_{\text{h1}} b_{\text{v2}}^2 b_{\text{v4}} & \frac{1}{2} i (-|0111\rangle
   -|1011\rangle +|1101\rangle +|1110\rangle ) & \frac{1}{512} \\
 b_{\text{h1}} b_{\text{v2}} b_{\text{v3}}^2 & \frac{1}{2} i (|0111\rangle
   -|1011\rangle -|1101\rangle +|1110\rangle ) & \frac{1}{512} \\
 b_{\text{h1}} b_{\text{v2}} b_{\text{v4}}^2 & -\frac{1}{2} i (|0111\rangle
   -|1011\rangle -|1101\rangle +|1110\rangle ) & \frac{1}{512} \\
 b_{\text{h1}} b_{\text{v3}}^2 b_{\text{v4}} & \frac{1}{2} i (|0111\rangle
   +|1011\rangle -|1101\rangle -|1110\rangle ) & \frac{1}{512} \\
 b_{\text{h1}} b_{\text{v3}} b_{\text{v4}}^2 & \frac{1}{2} (-|0111\rangle
   +|1011\rangle -|1101\rangle +|1110\rangle ) & \frac{1}{512} \\
 b_{\text{h2}}^2 b_{\text{h3}}^2 & |0000\rangle  & \frac{1}{256} \\
 b_{\text{h2}}^2 b_{\text{h3}} b_{\text{v1}} & \frac{1}{2} (|0001\rangle
   -|0010\rangle +|0100\rangle -|1000\rangle ) & \frac{1}{512} \\
 b_{\text{h2}}^2 b_{\text{h3}} b_{\text{v2}} & -\frac{1}{2} i (|0001\rangle
   +|0010\rangle -|0100\rangle -|1000\rangle ) & \frac{1}{512} \\
 b_{\text{h2}}^2 b_{\text{h3}} b_{\text{v3}} & \frac{1}{2} (|0001\rangle
   +|0010\rangle +|0100\rangle +|1000\rangle ) & \frac{1}{512} \\
 b_{\text{h2}}^2 b_{\text{h3}} b_{\text{v4}} & \frac{1}{2} i (|0001\rangle
   -|0010\rangle -|0100\rangle +|1000\rangle ) & \frac{1}{512} \\
 b_{\text{h2}}^2 b_{\text{h4}}^2 & -|0000\rangle  & \frac{1}{256} \\
 b_{\text{h2}}^2 b_{\text{h4}} b_{\text{v1}} & \frac{1}{2} i (|0001\rangle
   +|0010\rangle -|0100\rangle -|1000\rangle ) & \frac{1}{512} \\
 b_{\text{h2}}^2 b_{\text{h4}} b_{\text{v2}} & \frac{1}{2} (|0001\rangle
   -|0010\rangle +|0100\rangle -|1000\rangle ) & \frac{1}{512} \\
 b_{\text{h2}}^2 b_{\text{h4}} b_{\text{v3}} & \frac{1}{2} i (|0001\rangle
   -|0010\rangle -|0100\rangle +|1000\rangle ) & \frac{1}{512} \\
 b_{\text{h2}}^2 b_{\text{h4}} b_{\text{v4}} & \frac{1}{2} (-|0001\rangle
   -|0010\rangle -|0100\rangle -|1000\rangle ) & \frac{1}{512} \\
 b_{\text{h2}}^2 b_{\text{v1}}^2 & \frac{|0011\rangle +|0101\rangle
   -|0110\rangle -|1001\rangle +|1010\rangle +|1100\rangle }{\sqrt{6}} &
   \frac{3}{512} \\
 b_{\text{h2}}^2 b_{\text{v1}} b_{\text{v2}} & -\frac{i (|0110\rangle
   -|1001\rangle )}{\sqrt{2}} & \frac{1}{256} \\
 b_{\text{h2}}^2 b_{\text{v1}} b_{\text{v3}} & \frac{|0101\rangle -|1010\rangle
   }{\sqrt{2}} & \frac{1}{256} \\
 b_{\text{h2}}^2 b_{\text{v1}} b_{\text{v4}} & \frac{i (|0011\rangle
   -|1100\rangle )}{\sqrt{2}} & \frac{1}{256} \\
 b_{\text{h2}}^2 b_{\text{v2}}^2 & \frac{|0011\rangle +|0101\rangle
   +|0110\rangle +|1001\rangle +|1010\rangle +|1100\rangle }{\sqrt{6}} &
   \frac{3}{512} \\
 b_{\text{h2}}^2 b_{\text{v2}} b_{\text{v3}} & \frac{i (|0011\rangle
   -|1100\rangle )}{\sqrt{2}} & \frac{1}{256} \\
 b_{\text{h2}}^2 b_{\text{v2}} b_{\text{v4}} & \frac{|1010\rangle -|0101\rangle
   }{\sqrt{2}} & \frac{1}{256} \\
 b_{\text{h2}}^2 b_{\text{v3}}^2 & \frac{-|0011\rangle +|0101\rangle
   +|0110\rangle +|1001\rangle +|1010\rangle -|1100\rangle }{\sqrt{6}} &
   \frac{3}{512} \\
 b_{\text{h2}}^2 b_{\text{v3}} b_{\text{v4}} & -\frac{i (|0110\rangle
   -|1001\rangle )}{\sqrt{2}} & \frac{1}{256} \\
 b_{\text{h2}}^2 b_{\text{v4}}^2 & \frac{-|0011\rangle +|0101\rangle
   -|0110\rangle -|1001\rangle +|1010\rangle -|1100\rangle }{\sqrt{6}} &
   \frac{3}{512} \\
 b_{\text{h2}} b_{\text{h3}}^2 b_{\text{v1}} & \frac{1}{2} i (|0001\rangle
   -|0010\rangle -|0100\rangle +|1000\rangle ) & \frac{1}{512} \\
 b_{\text{h2}} b_{\text{h3}}^2 b_{\text{v2}} & \frac{1}{2} (|0001\rangle
   +|0010\rangle +|0100\rangle +|1000\rangle ) & \frac{1}{512} \\
 b_{\text{h2}} b_{\text{h3}}^2 b_{\text{v3}} & \frac{1}{2} i (|0001\rangle
   +|0010\rangle -|0100\rangle -|1000\rangle ) & \frac{1}{512} \\
 b_{\text{h2}} b_{\text{h3}}^2 b_{\text{v4}} & \frac{1}{2} (-|0001\rangle
   +|0010\rangle -|0100\rangle +|1000\rangle ) & \frac{1}{512} \\
 b_{\text{h2}} b_{\text{h3}} b_{\text{v1}}^2 & -\frac{i (|0011\rangle
   -|1100\rangle )}{\sqrt{2}} & \frac{1}{256} \\
 b_{\text{h2}} b_{\text{h3}} b_{\text{v2}}^2 & -\frac{i (|0011\rangle
   -|1100\rangle )}{\sqrt{2}} & \frac{1}{256} \\
 b_{\text{h2}} b_{\text{h3}} b_{\text{v3}}^2 & \frac{i (|0011\rangle
   -|1100\rangle )}{\sqrt{2}} & \frac{1}{256} \\
 b_{\text{h2}} b_{\text{h3}} b_{\text{v4}}^2 & \frac{i (|0011\rangle
   -|1100\rangle )}{\sqrt{2}} & \frac{1}{256} \\
 b_{\text{h2}} b_{\text{h4}}^2 b_{\text{v1}} & -\frac{1}{2} i (|0001\rangle
   -|0010\rangle -|0100\rangle +|1000\rangle ) & \frac{1}{512} \\
 b_{\text{h2}} b_{\text{h4}}^2 b_{\text{v2}} & \frac{1}{2} (-|0001\rangle
   -|0010\rangle -|0100\rangle -|1000\rangle ) & \frac{1}{512} \\
 b_{\text{h2}} b_{\text{h4}}^2 b_{\text{v3}} & -\frac{1}{2} i (|0001\rangle
   +|0010\rangle -|0100\rangle -|1000\rangle ) & \frac{1}{512} \\
 b_{\text{h2}} b_{\text{h4}}^2 b_{\text{v4}} & \frac{1}{2} (|0001\rangle
   -|0010\rangle +|0100\rangle -|1000\rangle ) & \frac{1}{512} \\
 b_{\text{h2}} b_{\text{h4}} b_{\text{v1}}^2 & \frac{|0101\rangle -|1010\rangle
   }{\sqrt{2}} & \frac{1}{256} \\
 b_{\text{h2}} b_{\text{h4}} b_{\text{v2}}^2 & \frac{|0101\rangle -|1010\rangle
   }{\sqrt{2}} & \frac{1}{256} \\
 b_{\text{h2}} b_{\text{h4}} b_{\text{v3}}^2 & \frac{|0101\rangle -|1010\rangle
   }{\sqrt{2}} & \frac{1}{256} \\
 b_{\text{h2}} b_{\text{h4}} b_{\text{v4}}^2 & \frac{|0101\rangle -|1010\rangle
   }{\sqrt{2}} & \frac{1}{256} \\
 b_{\text{h2}} b_{\text{v1}}^2 b_{\text{v2}} & \frac{1}{2} (|0111\rangle
   +|1011\rangle +|1101\rangle +|1110\rangle ) & \frac{1}{512} \\
 b_{\text{h2}} b_{\text{v1}}^2 b_{\text{v3}} & \frac{1}{2} i (-|0111\rangle
   -|1011\rangle +|1101\rangle +|1110\rangle ) & \frac{1}{512} \\
 b_{\text{h2}} b_{\text{v1}}^2 b_{\text{v4}} & \frac{1}{2} (|0111\rangle
   -|1011\rangle +|1101\rangle -|1110\rangle ) & \frac{1}{512} \\
 b_{\text{h2}} b_{\text{v1}} b_{\text{v2}}^2 & -\frac{1}{2} i (|0111\rangle
   -|1011\rangle -|1101\rangle +|1110\rangle ) & \frac{1}{512} \\
 b_{\text{h2}} b_{\text{v1}} b_{\text{v3}}^2 & \frac{1}{2} i (|0111\rangle
   -|1011\rangle -|1101\rangle +|1110\rangle ) & \frac{1}{512} \\
 b_{\text{h2}} b_{\text{v1}} b_{\text{v4}}^2 & -\frac{1}{2} i (|0111\rangle
   -|1011\rangle -|1101\rangle +|1110\rangle ) & \frac{1}{512} \\
 b_{\text{h2}} b_{\text{v2}}^2 b_{\text{v3}} & \frac{1}{2} i (|0111\rangle
   +|1011\rangle -|1101\rangle -|1110\rangle ) & \frac{1}{512} \\
 b_{\text{h2}} b_{\text{v2}}^2 b_{\text{v4}} & \frac{1}{2} (-|0111\rangle
   +|1011\rangle -|1101\rangle +|1110\rangle ) & \frac{1}{512} \\
 b_{\text{h2}} b_{\text{v2}} b_{\text{v3}}^2 & \frac{1}{2} (|0111\rangle
   +|1011\rangle +|1101\rangle +|1110\rangle ) & \frac{1}{512} \\
 b_{\text{h2}} b_{\text{v2}} b_{\text{v4}}^2 & \frac{1}{2} (-|0111\rangle
   -|1011\rangle -|1101\rangle -|1110\rangle ) & \frac{1}{512} \\
 b_{\text{h2}} b_{\text{v3}}^2 b_{\text{v4}} & \frac{1}{2} (|0111\rangle
   -|1011\rangle +|1101\rangle -|1110\rangle ) & \frac{1}{512} \\
 b_{\text{h2}} b_{\text{v3}} b_{\text{v4}}^2 & \frac{1}{2} i (|0111\rangle
   +|1011\rangle -|1101\rangle -|1110\rangle ) & \frac{1}{512} \\
 b_{\text{h3}}^2 b_{\text{h4}}^2 & |0000\rangle  & \frac{1}{256} \\
 b_{\text{h3}}^2 b_{\text{h4}} b_{\text{v1}} & -\frac{1}{2} i (|0001\rangle
   +|0010\rangle -|0100\rangle -|1000\rangle ) & \frac{1}{512} \\
 b_{\text{h3}}^2 b_{\text{h4}} b_{\text{v2}} & \frac{1}{2} (-|0001\rangle
   +|0010\rangle -|0100\rangle +|1000\rangle ) & \frac{1}{512} \\
 b_{\text{h3}}^2 b_{\text{h4}} b_{\text{v3}} & -\frac{1}{2} i (|0001\rangle
   -|0010\rangle -|0100\rangle +|1000\rangle ) & \frac{1}{512} \\
 b_{\text{h3}}^2 b_{\text{h4}} b_{\text{v4}} & \frac{1}{2} (|0001\rangle
   +|0010\rangle +|0100\rangle +|1000\rangle ) & \frac{1}{512} \\
 b_{\text{h3}}^2 b_{\text{v1}}^2 & \frac{-|0011\rangle +|0101\rangle
   -|0110\rangle -|1001\rangle +|1010\rangle -|1100\rangle }{\sqrt{6}} &
   \frac{3}{512} \\
 b_{\text{h3}}^2 b_{\text{v1}} b_{\text{v2}} & -\frac{i (|0110\rangle
   -|1001\rangle )}{\sqrt{2}} & \frac{1}{256} \\
 b_{\text{h3}}^2 b_{\text{v1}} b_{\text{v3}} & \frac{|0101\rangle -|1010\rangle
   }{\sqrt{2}} & \frac{1}{256} \\
 b_{\text{h3}}^2 b_{\text{v1}} b_{\text{v4}} & -\frac{i (|0011\rangle
   -|1100\rangle )}{\sqrt{2}} & \frac{1}{256} \\
 b_{\text{h3}}^2 b_{\text{v2}}^2 & \frac{-|0011\rangle +|0101\rangle
   +|0110\rangle +|1001\rangle +|1010\rangle -|1100\rangle }{\sqrt{6}} &
   \frac{3}{512} \\
 b_{\text{h3}}^2 b_{\text{v2}} b_{\text{v3}} & -\frac{i (|0011\rangle
   -|1100\rangle )}{\sqrt{2}} & \frac{1}{256} \\
 b_{\text{h3}}^2 b_{\text{v2}} b_{\text{v4}} & \frac{|1010\rangle -|0101\rangle
   }{\sqrt{2}} & \frac{1}{256} \\
 b_{\text{h3}}^2 b_{\text{v3}}^2 & \frac{|0011\rangle +|0101\rangle
   +|0110\rangle +|1001\rangle +|1010\rangle +|1100\rangle }{\sqrt{6}} &
   \frac{3}{512} \\
 b_{\text{h3}}^2 b_{\text{v3}} b_{\text{v4}} & -\frac{i (|0110\rangle
   -|1001\rangle )}{\sqrt{2}} & \frac{1}{256} \\
 b_{\text{h3}}^2 b_{\text{v4}}^2 & \frac{|0011\rangle +|0101\rangle
   -|0110\rangle -|1001\rangle +|1010\rangle +|1100\rangle }{\sqrt{6}} &
   \frac{3}{512} \\
 b_{\text{h3}} b_{\text{h4}}^2 b_{\text{v1}} & \frac{1}{2} (|0001\rangle
   -|0010\rangle +|0100\rangle -|1000\rangle ) & \frac{1}{512} \\
 b_{\text{h3}} b_{\text{h4}}^2 b_{\text{v2}} & -\frac{1}{2} i (|0001\rangle
   +|0010\rangle -|0100\rangle -|1000\rangle ) & \frac{1}{512} \\
 b_{\text{h3}} b_{\text{h4}}^2 b_{\text{v3}} & \frac{1}{2} (|0001\rangle
   +|0010\rangle +|0100\rangle +|1000\rangle ) & \frac{1}{512} \\
 b_{\text{h3}} b_{\text{h4}}^2 b_{\text{v4}} & \frac{1}{2} i (|0001\rangle
   -|0010\rangle -|0100\rangle +|1000\rangle ) & \frac{1}{512} \\
 b_{\text{h3}} b_{\text{h4}} b_{\text{v1}}^2 & -\frac{i (|0110\rangle
   -|1001\rangle )}{\sqrt{2}} & \frac{1}{256} \\
 b_{\text{h3}} b_{\text{h4}} b_{\text{v2}}^2 & \frac{i (|0110\rangle
   -|1001\rangle )}{\sqrt{2}} & \frac{1}{256} \\
 b_{\text{h3}} b_{\text{h4}} b_{\text{v3}}^2 & \frac{i (|0110\rangle
   -|1001\rangle )}{\sqrt{2}} & \frac{1}{256} \\
 b_{\text{h3}} b_{\text{h4}} b_{\text{v4}}^2 & -\frac{i (|0110\rangle
   -|1001\rangle )}{\sqrt{2}} & \frac{1}{256} \\
 b_{\text{h3}} b_{\text{v1}}^2 b_{\text{v2}} & \frac{1}{2} i (-|0111\rangle
   -|1011\rangle +|1101\rangle +|1110\rangle ) & \frac{1}{512} \\
 b_{\text{h3}} b_{\text{v1}}^2 b_{\text{v3}} & \frac{1}{2} (-|0111\rangle
   -|1011\rangle -|1101\rangle -|1110\rangle ) & \frac{1}{512} \\
 b_{\text{h3}} b_{\text{v1}}^2 b_{\text{v4}} & -\frac{1}{2} i (|0111\rangle
   -|1011\rangle -|1101\rangle +|1110\rangle ) & \frac{1}{512} \\
 b_{\text{h3}} b_{\text{v1}} b_{\text{v2}}^2 & \frac{1}{2} (-|0111\rangle
   +|1011\rangle -|1101\rangle +|1110\rangle ) & \frac{1}{512} \\
 b_{\text{h3}} b_{\text{v1}} b_{\text{v3}}^2 & \frac{1}{2} (|0111\rangle
   -|1011\rangle +|1101\rangle -|1110\rangle ) & \frac{1}{512} \\
 b_{\text{h3}} b_{\text{v1}} b_{\text{v4}}^2 & \frac{1}{2} (-|0111\rangle
   +|1011\rangle -|1101\rangle +|1110\rangle ) & \frac{1}{512} \\
 b_{\text{h3}} b_{\text{v2}}^2 b_{\text{v3}} & \frac{1}{2} (|0111\rangle
   +|1011\rangle +|1101\rangle +|1110\rangle ) & \frac{1}{512} \\
 b_{\text{h3}} b_{\text{v2}}^2 b_{\text{v4}} & \frac{1}{2} i (|0111\rangle
   -|1011\rangle -|1101\rangle +|1110\rangle ) & \frac{1}{512} \\
 b_{\text{h3}} b_{\text{v2}} b_{\text{v3}}^2 & \frac{1}{2} i (-|0111\rangle
   -|1011\rangle +|1101\rangle +|1110\rangle ) & \frac{1}{512} \\
 b_{\text{h3}} b_{\text{v2}} b_{\text{v4}}^2 & \frac{1}{2} i (|0111\rangle
   +|1011\rangle -|1101\rangle -|1110\rangle ) & \frac{1}{512} \\
 b_{\text{h3}} b_{\text{v3}}^2 b_{\text{v4}} & -\frac{1}{2} i (|0111\rangle
   -|1011\rangle -|1101\rangle +|1110\rangle ) & \frac{1}{512} \\
 b_{\text{h3}} b_{\text{v3}} b_{\text{v4}}^2 & \frac{1}{2} (|0111\rangle
   +|1011\rangle +|1101\rangle +|1110\rangle ) & \frac{1}{512} \\
 b_{\text{h4}}^2 b_{\text{v1}}^2 & \frac{-|0011\rangle +|0101\rangle
   +|0110\rangle +|1001\rangle +|1010\rangle -|1100\rangle }{\sqrt{6}} &
   \frac{3}{512} \\
 b_{\text{h4}}^2 b_{\text{v1}} b_{\text{v2}} & \frac{i (|0110\rangle
   -|1001\rangle )}{\sqrt{2}} & \frac{1}{256} \\
 b_{\text{h4}}^2 b_{\text{v1}} b_{\text{v3}} & \frac{|0101\rangle -|1010\rangle
   }{\sqrt{2}} & \frac{1}{256} \\
 b_{\text{h4}}^2 b_{\text{v1}} b_{\text{v4}} & -\frac{i (|0011\rangle
   -|1100\rangle )}{\sqrt{2}} & \frac{1}{256} \\
 b_{\text{h4}}^2 b_{\text{v2}}^2 & \frac{-|0011\rangle +|0101\rangle
   -|0110\rangle -|1001\rangle +|1010\rangle -|1100\rangle }{\sqrt{6}} &
   \frac{3}{512} \\
 b_{\text{h4}}^2 b_{\text{v2}} b_{\text{v3}} & -\frac{i (|0011\rangle
   -|1100\rangle )}{\sqrt{2}} & \frac{1}{256} \\
 b_{\text{h4}}^2 b_{\text{v2}} b_{\text{v4}} & \frac{|1010\rangle -|0101\rangle
   }{\sqrt{2}} & \frac{1}{256} \\
 b_{\text{h4}}^2 b_{\text{v3}}^2 & \frac{|0011\rangle +|0101\rangle
   -|0110\rangle -|1001\rangle +|1010\rangle +|1100\rangle }{\sqrt{6}} &
   \frac{3}{512} \\
 b_{\text{h4}}^2 b_{\text{v3}} b_{\text{v4}} & \frac{i (|0110\rangle
   -|1001\rangle )}{\sqrt{2}} & \frac{1}{256} \\
 b_{\text{h4}}^2 b_{\text{v4}}^2 & \frac{|0011\rangle +|0101\rangle
   +|0110\rangle +|1001\rangle +|1010\rangle +|1100\rangle }{\sqrt{6}} &
   \frac{3}{512} \\
 b_{\text{h4}} b_{\text{v1}}^2 b_{\text{v2}} & \frac{1}{2} (|0111\rangle
   -|1011\rangle +|1101\rangle -|1110\rangle ) & \frac{1}{512} \\
 b_{\text{h4}} b_{\text{v1}}^2 b_{\text{v3}} & -\frac{1}{2} i (|0111\rangle
   -|1011\rangle -|1101\rangle +|1110\rangle ) & \frac{1}{512} \\
 b_{\text{h4}} b_{\text{v1}}^2 b_{\text{v4}} & \frac{1}{2} (|0111\rangle
   +|1011\rangle +|1101\rangle +|1110\rangle ) & \frac{1}{512} \\
 b_{\text{h4}} b_{\text{v1}} b_{\text{v2}}^2 & \frac{1}{2} i (-|0111\rangle
   -|1011\rangle +|1101\rangle +|1110\rangle ) & \frac{1}{512} \\
 b_{\text{h4}} b_{\text{v1}} b_{\text{v3}}^2 & \frac{1}{2} i (|0111\rangle
   +|1011\rangle -|1101\rangle -|1110\rangle ) & \frac{1}{512} \\
 b_{\text{h4}} b_{\text{v1}} b_{\text{v4}}^2 & \frac{1}{2} i (-|0111\rangle
   -|1011\rangle +|1101\rangle +|1110\rangle ) & \frac{1}{512} \\
 b_{\text{h4}} b_{\text{v2}}^2 b_{\text{v3}} & \frac{1}{2} i (|0111\rangle
   -|1011\rangle -|1101\rangle +|1110\rangle ) & \frac{1}{512} \\
 b_{\text{h4}} b_{\text{v2}}^2 b_{\text{v4}} & \frac{1}{2} (-|0111\rangle
   -|1011\rangle -|1101\rangle -|1110\rangle ) & \frac{1}{512} \\
 b_{\text{h4}} b_{\text{v2}} b_{\text{v3}}^2 & \frac{1}{2} (|0111\rangle
   -|1011\rangle +|1101\rangle -|1110\rangle ) & \frac{1}{512} \\
 b_{\text{h4}} b_{\text{v2}} b_{\text{v4}}^2 & \frac{1}{2} (-|0111\rangle
   +|1011\rangle -|1101\rangle +|1110\rangle ) & \frac{1}{512} \\
 b_{\text{h4}} b_{\text{v3}}^2 b_{\text{v4}} & \frac{1}{2} (|0111\rangle
   +|1011\rangle +|1101\rangle +|1110\rangle ) & \frac{1}{512} \\
 b_{\text{h4}} b_{\text{v3}} b_{\text{v4}}^2 & \frac{1}{2} i (|0111\rangle
   -|1011\rangle -|1101\rangle +|1110\rangle ) & \frac{1}{512} \\
 b_{\text{v1}}^2 b_{\text{v2}}^2 & |1111\rangle  & \frac{1}{256} \\
 b_{\text{v1}}^2 b_{\text{v3}}^2 & -|1111\rangle  & \frac{1}{256} \\
 b_{\text{v1}}^2 b_{\text{v4}}^2 & |1111\rangle  & \frac{1}{256} \\
 b_{\text{v2}}^2 b_{\text{v3}}^2 & |1111\rangle  & \frac{1}{256} \\
 b_{\text{v2}}^2 b_{\text{v4}}^2 & -|1111\rangle  & \frac{1}{256} \\
 b_{\text{v3}}^2 b_{\text{v4}}^2 & |1111\rangle  & \frac{1}{256} 
\end{eqnarray*}

If we consider at most three clicks per detector, the possible patterns are:
\begin{eqnarray*}
    \text{\underline{Detection Pattern}} & \text{\underline{State}} & \text{\underline{Probability}} \\
 b_{\text{h1}}^3 b_{\text{v1}} & \frac{1}{2} (|0001\rangle +|0010\rangle
   +|0100\rangle +|1000\rangle ) & \frac{3}{512} \\
 b_{\text{h1}}^3 b_{\text{v2}} & -\frac{1}{2} i (|0001\rangle -|0010\rangle
   -|0100\rangle +|1000\rangle ) & \frac{3}{512} \\
 b_{\text{h1}}^3 b_{\text{v3}} & \frac{1}{2} (|0001\rangle -|0010\rangle
   +|0100\rangle -|1000\rangle ) & \frac{3}{512} \\
 b_{\text{h1}}^3 b_{\text{v4}} & \frac{1}{2} i (|0001\rangle +|0010\rangle
   -|0100\rangle -|1000\rangle ) & \frac{3}{512} \\
 b_{\text{h1}} b_{\text{v1}}^3 & \frac{1}{2} (|0111\rangle +|1011\rangle
   +|1101\rangle +|1110\rangle ) & \frac{3}{512} \\
 b_{\text{h1}} b_{\text{v2}}^3 & \frac{1}{2} i (|0111\rangle -|1011\rangle
   -|1101\rangle +|1110\rangle ) & \frac{3}{512} \\
 b_{\text{h1}} b_{\text{v3}}^3 & \frac{1}{2} (-|0111\rangle +|1011\rangle
   -|1101\rangle +|1110\rangle ) & \frac{3}{512} \\
 b_{\text{h1}} b_{\text{v4}}^3 & \frac{1}{2} i (|0111\rangle +|1011\rangle
   -|1101\rangle -|1110\rangle ) & \frac{3}{512} \\
 b_{\text{h2}}^3 b_{\text{v1}} & \frac{1}{2} i (|0001\rangle -|0010\rangle
   -|0100\rangle +|1000\rangle ) & \frac{3}{512} \\
 b_{\text{h2}}^3 b_{\text{v2}} & \frac{1}{2} (|0001\rangle +|0010\rangle
   +|0100\rangle +|1000\rangle ) & \frac{3}{512} \\
 b_{\text{h2}}^3 b_{\text{v3}} & \frac{1}{2} i (|0001\rangle +|0010\rangle
   -|0100\rangle -|1000\rangle ) & \frac{3}{512} \\
 b_{\text{h2}}^3 b_{\text{v4}} & \frac{1}{2} (-|0001\rangle +|0010\rangle
   -|0100\rangle +|1000\rangle ) & \frac{3}{512} \\
 b_{\text{h2}} b_{\text{v1}}^3 & -\frac{1}{2} i (|0111\rangle -|1011\rangle
   -|1101\rangle +|1110\rangle ) & \frac{3}{512} \\
 b_{\text{h2}} b_{\text{v2}}^3 & \frac{1}{2} (|0111\rangle +|1011\rangle
   +|1101\rangle +|1110\rangle ) & \frac{3}{512} \\
 b_{\text{h2}} b_{\text{v3}}^3 & \frac{1}{2} i (|0111\rangle +|1011\rangle
   -|1101\rangle -|1110\rangle ) & \frac{3}{512} \\
 b_{\text{h2}} b_{\text{v4}}^3 & \frac{1}{2} (|0111\rangle -|1011\rangle
   +|1101\rangle -|1110\rangle ) & \frac{3}{512} \\
 b_{\text{h3}}^3 b_{\text{v1}} & \frac{1}{2} (|0001\rangle -|0010\rangle
   +|0100\rangle -|1000\rangle ) & \frac{3}{512} \\
 b_{\text{h3}}^3 b_{\text{v2}} & -\frac{1}{2} i (|0001\rangle +|0010\rangle
   -|0100\rangle -|1000\rangle ) & \frac{3}{512} \\
 b_{\text{h3}}^3 b_{\text{v3}} & \frac{1}{2} (|0001\rangle +|0010\rangle
   +|0100\rangle +|1000\rangle ) & \frac{3}{512} \\
 b_{\text{h3}}^3 b_{\text{v4}} & \frac{1}{2} i (|0001\rangle -|0010\rangle
   -|0100\rangle +|1000\rangle ) & \frac{3}{512} \\
 b_{\text{h3}} b_{\text{v1}}^3 & \frac{1}{2} (-|0111\rangle +|1011\rangle
   -|1101\rangle +|1110\rangle ) & \frac{3}{512} \\
 b_{\text{h3}} b_{\text{v2}}^3 & \frac{1}{2} i (-|0111\rangle -|1011\rangle
   +|1101\rangle +|1110\rangle ) & \frac{3}{512} \\
 b_{\text{h3}} b_{\text{v3}}^3 & \frac{1}{2} (|0111\rangle +|1011\rangle
   +|1101\rangle +|1110\rangle ) & \frac{3}{512} \\
 b_{\text{h3}} b_{\text{v4}}^3 & -\frac{1}{2} i (|0111\rangle -|1011\rangle
   -|1101\rangle +|1110\rangle ) & \frac{3}{512} \\
 b_{\text{h4}}^3 b_{\text{v1}} & -\frac{1}{2} i (|0001\rangle +|0010\rangle
   -|0100\rangle -|1000\rangle ) & \frac{3}{512} \\
 b_{\text{h4}}^3 b_{\text{v2}} & \frac{1}{2} (-|0001\rangle +|0010\rangle
   -|0100\rangle +|1000\rangle ) & \frac{3}{512} \\
 b_{\text{h4}}^3 b_{\text{v3}} & -\frac{1}{2} i (|0001\rangle -|0010\rangle
   -|0100\rangle +|1000\rangle ) & \frac{3}{512} \\
 b_{\text{h4}}^3 b_{\text{v4}} & \frac{1}{2} (|0001\rangle +|0010\rangle
   +|0100\rangle +|1000\rangle ) & \frac{3}{512} \\
 b_{\text{h4}} b_{\text{v1}}^3 & \frac{1}{2} i (-|0111\rangle -|1011\rangle
   +|1101\rangle +|1110\rangle ) & \frac{3}{512} \\
 b_{\text{h4}} b_{\text{v2}}^3 & \frac{1}{2} (|0111\rangle -|1011\rangle
   +|1101\rangle -|1110\rangle ) & \frac{3}{512} \\
 b_{\text{h4}} b_{\text{v3}}^3 & \frac{1}{2} i (|0111\rangle -|1011\rangle
   -|1101\rangle +|1110\rangle ) & \frac{3}{512} \\
 b_{\text{h4}} b_{\text{v4}}^3 & \frac{1}{2} (|0111\rangle +|1011\rangle
   +|1101\rangle +|1110\rangle ) & \frac{3}{512} \\
\end{eqnarray*}

If we consider four clicks per detector, the possible patterns are:
\begin{eqnarray*}
 \text{\underline{Detection Pattern}} & \text{\underline{State}} & \text{\underline{Probability}} \\
 b_{\text{h1}}^4 & |0000\rangle  & \frac{3}{512} \\
 b_{\text{h2}}^4 & |0000\rangle  & \frac{3}{512} \\
 b_{\text{h3}}^4 & |0000\rangle  & \frac{3}{512} \\
 b_{\text{h4}}^4 & |0000\rangle  & \frac{3}{512} \\
 b_{\text{v1}}^4 & |1111\rangle  & \frac{3}{512} \\
 b_{\text{v2}}^4 & |1111\rangle  & \frac{3}{512} \\
 b_{\text{v3}}^4 & |1111\rangle  & \frac{3}{512} \\
 b_{\text{v4}}^4 & |1111\rangle  & \frac{3}{512}.
\end{eqnarray*}

Notice that not all detections lead to entangled atomic states. From this table we can see that, if we have non-number-resolved photodetectors, and we condition on the detection of four photons in 4 different detectors, the probability of generating entanglement between four atoms is proportional to $(7/32)\eta^4$. If number-resolved detectors are available, and we admit any combination of detections, the probability increases to to $(7/8)\eta^4$.

This particular implementation of quarter exhibit Hong-Ou-Mandel suppression of the following detection combinations:

\begin{eqnarray*}
 b_{\text{h1}}^3 b_{\text{h2}} \\
 b_{\text{h1}}^3 b_{\text{h3}} \\
 b_{\text{h1}}^3 b_{\text{h4}} \\
 b_{\text{h1}}^2 b_{\text{h2}} b_{\text{h3}} \\
 b_{\text{h1}}^2 b_{\text{h2}} b_{\text{h4}} \\
 b_{\text{h1}}^2 b_{\text{h3}} b_{\text{h4}} \\
 b_{\text{h1}} b_{\text{h2}}^3 \\
 b_{\text{h1}} b_{\text{h2}}^2 b_{\text{h3}} \\
 b_{\text{h1}} b_{\text{h2}}^2 b_{\text{h4}} \\
 b_{\text{h1}} b_{\text{h2}} b_{\text{h3}}^2 \\
 b_{\text{h1}} b_{\text{h2}} b_{\text{h4}}^2 \\
 b_{\text{h1}} b_{\text{h2}} b_{\text{v1}} b_{\text{v3}} \\
 b_{\text{h1}} b_{\text{h2}} b_{\text{v1}} b_{\text{v4}} \\
 b_{\text{h1}} b_{\text{h2}} b_{\text{v2}} b_{\text{v3}} \\
 b_{\text{h1}} b_{\text{h2}} b_{\text{v2}} b_{\text{v4}} \\
 b_{\text{h1}} b_{\text{h3}}^3 \\
 b_{\text{h1}} b_{\text{h3}}^2 b_{\text{h4}} \\
 b_{\text{h1}} b_{\text{h3}} b_{\text{h4}}^2 \\
 b_{\text{h1}} b_{\text{h3}} b_{\text{v1}} b_{\text{v2}} \\
 b_{\text{h1}} b_{\text{h3}} b_{\text{v1}} b_{\text{v4}} \\
 b_{\text{h1}} b_{\text{h3}} b_{\text{v2}} b_{\text{v3}} \\
 b_{\text{h1}} b_{\text{h3}} b_{\text{v3}} b_{\text{v4}} \\
 b_{\text{h1}} b_{\text{h4}}^3 \\
 b_{\text{h1}} b_{\text{h4}} b_{\text{v1}} b_{\text{v2}} \\
 b_{\text{h1}} b_{\text{h4}} b_{\text{v1}} b_{\text{v3}} \\
 b_{\text{h1}} b_{\text{h4}} b_{\text{v2}} b_{\text{v4}} \\
 b_{\text{h1}} b_{\text{h4}} b_{\text{v3}} b_{\text{v4}} \\
 b_{\text{h2}}^3 b_{\text{h3}} \\
 b_{\text{h2}}^3 b_{\text{h4}} \\
 b_{\text{h2}}^2 b_{\text{h3}} b_{\text{h4}} \\
 b_{\text{h2}} b_{\text{h3}}^3 \\
 b_{\text{h2}} b_{\text{h3}}^2 b_{\text{h4}} \\
 b_{\text{h2}} b_{\text{h3}} b_{\text{h4}}^2 \\
 b_{\text{h2}} b_{\text{h3}} b_{\text{v1}} b_{\text{v2}} \\
 b_{\text{h2}} b_{\text{h3}} b_{\text{v1}} b_{\text{v3}} \\
 b_{\text{h2}} b_{\text{h3}} b_{\text{v2}} b_{\text{v4}} \\
 b_{\text{h2}} b_{\text{h3}} b_{\text{v3}} b_{\text{v4}} \\
 b_{\text{h2}} b_{\text{h4}}^3 \\
 b_{\text{h2}} b_{\text{h4}} b_{\text{v1}} b_{\text{v2}} \\
 b_{\text{h2}} b_{\text{h4}} b_{\text{v1}} b_{\text{v4}} \\
 b_{\text{h2}} b_{\text{h4}} b_{\text{v2}} b_{\text{v3}} \\
 b_{\text{h2}} b_{\text{h4}} b_{\text{v3}} b_{\text{v4}} \\
 b_{\text{h3}}^3 b_{\text{h4}} \\
 b_{\text{h3}} b_{\text{h4}}^3 \\
 b_{\text{h3}} b_{\text{h4}} b_{\text{v1}} b_{\text{v3}} \\
 b_{\text{h3}} b_{\text{h4}} b_{\text{v1}} b_{\text{v4}} \\
 b_{\text{h3}} b_{\text{h4}} b_{\text{v2}} b_{\text{v3}} \\
 b_{\text{h3}} b_{\text{h4}} b_{\text{v2}} b_{\text{v4}} \\
 b_{\text{v1}}^3 b_{\text{v2}} \\
 b_{\text{v1}}^3 b_{\text{v3}} \\
 b_{\text{v1}}^3 b_{\text{v4}} \\
 b_{\text{v1}}^2 b_{\text{v2}} b_{\text{v3}} \\
 b_{\text{v1}}^2 b_{\text{v2}} b_{\text{v4}} \\
 b_{\text{v1}}^2 b_{\text{v3}} b_{\text{v4}} \\
 b_{\text{v1}} b_{\text{v2}}^3 \\
 b_{\text{v1}} b_{\text{v2}}^2 b_{\text{v3}} \\
 b_{\text{v1}} b_{\text{v2}}^2 b_{\text{v4}} \\
 b_{\text{v1}} b_{\text{v2}} b_{\text{v3}}^2 \\
 b_{\text{v1}} b_{\text{v2}} b_{\text{v4}}^2 \\
 b_{\text{v1}} b_{\text{v3}}^3 \\
 b_{\text{v1}} b_{\text{v3}}^2 b_{\text{v4}} \\
 b_{\text{v1}} b_{\text{v3}} b_{\text{v4}}^2 \\
 b_{\text{v1}} b_{\text{v4}}^3 \\
 b_{\text{v2}}^3 b_{\text{v3}} \\
 b_{\text{v2}}^3 b_{\text{v4}} \\
 b_{\text{v2}}^2 b_{\text{v3}} b_{\text{v4}} \\
 b_{\text{v2}} b_{\text{v3}}^3 \\
 b_{\text{v2}} b_{\text{v3}}^2 b_{\text{v4}} \\
 b_{\text{v2}} b_{\text{v3}} b_{\text{v4}}^2 \\
 b_{\text{v2}} b_{\text{v4}}^3 \\
 b_{\text{v3}}^3 b_{\text{v4}} \\
 b_{\text{v3}} b_{\text{v4}}^3 \\
\end{eqnarray*}

\onecolumngrid
%\appendix
\section{Entanglement swapping using a tritter}
\begin{figure}[b!]
\centerline{\includegraphics[width=0.3\columnwidth]{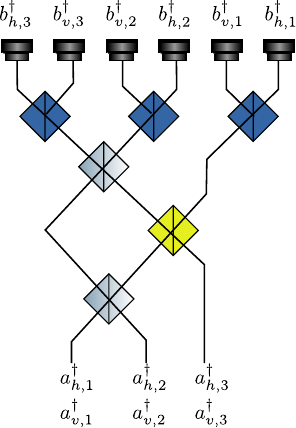}}
\caption{Input and output for a tritter.}
\label{fig:tritter}
\end{figure}

The same treatment can be applied for the case of 3-atom entanglement swapping  using polarisation encoded photons. In this case the transformation between input modes $a$ and output modes $b$ of the tritter is given by
\begin{align}
\left(
\begin{array}{c}
 b^\dagger_{\text{\{h,v\}1}} \\
 b^\dagger_{\text{\{h,v\}2}} \\
 b^\dagger_{\text{\{h,v\}3}} \\
\end{array}
\right)= T \cdot \left(
\begin{array}{c}
 a^\dagger_{\text{\{h,v\}1}} \\
 a^\dagger_{\text{\{h,v\}2}} \\
 a^\dagger_{\text{\{h,v\}3}} \\
\end{array}
\right)
\end{align}
The unitary transformation of the titter (see Fig.~\ref{fig:tritter}) is given by
\begin{align}
   T =  \left(
\begin{array}{ccc}
 1 & 0 & 0 \\
 0 & \frac{1}{\sqrt{2}} & \frac{i}{\sqrt{2}} \\
 0 & \frac{i}{\sqrt{2}} & \frac{1}{\sqrt{2}} \\
\end{array}
\right) \cdot \left(
\begin{array}{ccc}
 \sqrt{\frac{2}{3}} & 0 & \frac{i}{\sqrt{3}} \\
 0 & 1 & 0 \\
 \frac{i}{\sqrt{3}} & 0 & \sqrt{\frac{2}{3}} \\
\end{array}
\right)\cdot \left(
\begin{array}{ccc}
 \frac{1}{\sqrt{2}} & \frac{i}{\sqrt{2}} & 0 \\
 \frac{i}{\sqrt{2}} & \frac{1}{\sqrt{2}} & 0 \\
 0 & 0 & 1 \\
\end{array}
\right)
\end{align}
with the inverse matrix given by
\begin{align}
    T^{-1}=\left(
\begin{array}{ccc}
 \frac{1}{\sqrt{3}} & \frac{1}{6} \left(-\sqrt{3}-3 i\right) & -\frac{1}{6} i
   \left(\sqrt{3}-3 i\right) \\
 -\frac{i}{\sqrt{3}} & \frac{1}{6} \left(3+i \sqrt{3}\right) & \frac{1}{6}
   \left(-\sqrt{3}-3 i\right) \\
 -\frac{i}{\sqrt{3}} & -\frac{i}{\sqrt{3}} & \frac{1}{\sqrt{3}} \\
\end{array}
\right).
\end{align}

Following the same procedure as before, if we just consider the initial state to be
\begin{align}
    \bigotimes\limits_{n=1}^3 \frac{1}{\sqrt{2}}\left( \ket{0}_{A_n}\ket{0}_{P_n} + \ket{1}_{A_n}\ket{1}_{P_n} \right),
\end{align}
and allow for, at most, one click per detector, the possible detection patterns are
\begin{eqnarray*}
 \text{\underline{Detection Pattern}} & \text{\underline{State}} & \text{\underline{Probability}} \\
 b_{\text{h1}} b_{\text{h2}} b_{\text{h3}} & i |000\rangle  & \frac{1}{24} \\
 b_{\text{h1}} b_{\text{h2}} b_{\text{v1}} & \frac{2 \sqrt{3} |001\rangle -\left(\sqrt{3}-3
   i\right) |010\rangle }{2 \sqrt{6}} & \frac{1}{108} \\
 b_{\text{h1}} b_{\text{h2}} b_{\text{v2}} & \frac{2 \sqrt{3} |001\rangle -\left(\sqrt{3}+3
   i\right) |010\rangle }{2 \sqrt{6}} & \frac{1}{108} \\
 b_{\text{h1}} b_{\text{h2}} b_{\text{v3}} & \frac{i (|001\rangle +|010\rangle )}{\sqrt{2}} &
   \frac{1}{108} \\
 b_{\text{h1}} b_{\text{h3}} b_{\text{v1}} & \frac{2 i \sqrt{3} |001\rangle +\left(3-i
   \sqrt{3}\right) |010\rangle }{2 \sqrt{6}} & \frac{1}{108} \\
 b_{\text{h1}} b_{\text{h3}} b_{\text{v2}} & \frac{i (|001\rangle +|010\rangle )}{\sqrt{2}} &
   \frac{1}{108} \\
 b_{\text{h1}} b_{\text{h3}} b_{\text{v3}} & \frac{-2 \sqrt{3} |001\rangle +\left(\sqrt{3}-3
   i\right) |010\rangle }{2 \sqrt{6}} & \frac{1}{108} \\
 b_{\text{h1}} b_{\text{v1}} b_{\text{v2}} & \frac{-\left(\sqrt{3}+3 i\right) |011\rangle
   -\left(\sqrt{3}-3 i\right) |101\rangle +4 \sqrt{3} |110\rangle }{4 \sqrt{3}} & \frac{1}{54}
   \\
 b_{\text{h1}} b_{\text{v1}} b_{\text{v3}} & \frac{\left(-3-i \sqrt{3}\right) |011\rangle
   +\left(3-i \sqrt{3}\right) |101\rangle +4 i \sqrt{3} |110\rangle }{4 \sqrt{3}} & \frac{1}{54}
   \\
 b_{\text{h1}} b_{\text{v2}} b_{\text{v3}} & \frac{1}{2} i (|011\rangle +|101\rangle +2
   |110\rangle ) & \frac{1}{54} \\
 b_{\text{h2}} b_{\text{h3}} b_{\text{v1}} & \frac{i (|001\rangle +|010\rangle )}{\sqrt{2}} &
   \frac{1}{108} \\
 b_{\text{h2}} b_{\text{h3}} b_{\text{v2}} & \frac{2 i |001\rangle -\left(\sqrt{3}+i\right)
   |010\rangle }{2 \sqrt{2}} & \frac{1}{108} \\
 b_{\text{h2}} b_{\text{h3}} b_{\text{v3}} & \frac{-2 |001\rangle +\left(1+i \sqrt{3}\right)
   |010\rangle }{2 \sqrt{2}} & \frac{1}{108} \\
 b_{\text{h2}} b_{\text{v1}} b_{\text{v2}} & \frac{-\left(\sqrt{3}-3 i\right) |011\rangle
   -\left(\sqrt{3}+3 i\right) |101\rangle +4 \sqrt{3} |110\rangle }{4 \sqrt{3}} & \frac{1}{54}
   \\
 b_{\text{h2}} b_{\text{v1}} b_{\text{v3}} & \frac{1}{2} i (|011\rangle +|101\rangle +2
   |110\rangle ) & \frac{1}{54} \\
 b_{\text{h2}} b_{\text{v2}} b_{\text{v3}} & \frac{1}{4} \left(\left(\sqrt{3}-i\right)
   |011\rangle -\left(\sqrt{3}+i\right) |101\rangle +4 i |110\rangle \right) & \frac{1}{54} \\
 b_{\text{h3}} b_{\text{v1}} b_{\text{v2}} & \frac{1}{2} i (|011\rangle +|101\rangle +2
   |110\rangle ) & \frac{1}{54} \\
 b_{\text{h3}} b_{\text{v1}} b_{\text{v3}} & \frac{\left(\sqrt{3}+3 i\right) |011\rangle
   +\left(\sqrt{3}-3 i\right) |101\rangle -4 \sqrt{3} |110\rangle }{4 \sqrt{3}} & \frac{1}{54}
   \\
 b_{\text{h3}} b_{\text{v2}} b_{\text{v3}} & \frac{1}{4} \left(\left(1-i \sqrt{3}\right)
   |011\rangle +\left(1+i \sqrt{3}\right) |101\rangle -4 |110\rangle \right) & \frac{1}{54} \\
 b_{\text{v1}} b_{\text{v2}} b_{\text{v3}} & i |111\rangle  & \frac{1}{24} \\
\end{eqnarray*}

If two clicks per detector are allowed, then the possible detection combinations are
\begin{eqnarray*}
 \text{\underline{Detection Pattern}} & \text{\underline{State}} & \text{\underline{Probability}} \\
 b_{\text{h1}}^2 b_{\text{v1}} & -\frac{|001\rangle +|010\rangle }{\sqrt{2}} & \frac{1}{54} \\
 b_{\text{h1}}^2 b_{\text{v2}} & \frac{-2 \sqrt{3} |001\rangle +\left(\sqrt{3}-3 i\right)
   |010\rangle }{2 \sqrt{6}} & \frac{1}{54} \\
 b_{\text{h1}}^2 b_{\text{v3}} & \frac{i \left(\sqrt{3}+3 i\right) |010\rangle -2 i \sqrt{3}
   |001\rangle }{2 \sqrt{6}} & \frac{1}{54} \\
 b_{\text{h1}} b_{\text{v1}}^2 & \frac{1}{2} (-|011\rangle -|101\rangle -2 |110\rangle ) &
   \frac{1}{27} \\
 b_{\text{h1}} b_{\text{v2}}^2 & \frac{\left(\sqrt{3}-3 i\right) |011\rangle +\left(\sqrt{3}+3
   i\right) |101\rangle -4 \sqrt{3} |110\rangle }{4 \sqrt{3}} & \frac{1}{27} \\
 b_{\text{h1}} b_{\text{v3}}^2 & \frac{-\left(\sqrt{3}+3 i\right) |011\rangle -\left(\sqrt{3}-3
   i\right) |101\rangle +4 \sqrt{3} |110\rangle }{4 \sqrt{3}} & \frac{1}{27} \\
 b_{\text{h2}}^2 b_{\text{v1}} & \frac{-2 \sqrt{3} |001\rangle +\left(\sqrt{3}+3 i\right)
   |010\rangle }{2 \sqrt{6}} & \frac{1}{54} \\
 b_{\text{h2}}^2 b_{\text{v2}} & -\frac{|001\rangle +|010\rangle }{\sqrt{2}} & \frac{1}{54} \\
 b_{\text{h2}}^2 b_{\text{v3}} & \frac{\left(\sqrt{3}+i\right) |010\rangle -2 i |001\rangle }{2
   \sqrt{2}} & \frac{1}{54} \\
 b_{\text{h2}} b_{\text{v1}}^2 & \frac{\left(\sqrt{3}+3 i\right) |011\rangle +\left(\sqrt{3}-3
   i\right) |101\rangle -4 \sqrt{3} |110\rangle }{4 \sqrt{3}} & \frac{1}{27} \\
 b_{\text{h2}} b_{\text{v2}}^2 & \frac{1}{2} (-|011\rangle -|101\rangle -2 |110\rangle ) &
   \frac{1}{27} \\
 b_{\text{h2}} b_{\text{v3}}^2 & \frac{1}{4} \left(i \left(\sqrt{3}+i\right) |011\rangle
   +\left(-1-i \sqrt{3}\right) |101\rangle +4 |110\rangle \right) & \frac{1}{27} \\
 b_{\text{h3}}^2 b_{\text{v1}} & \frac{4 \sqrt{3} |001\rangle -2 \left(\sqrt{3}-3 i\right)
   |010\rangle }{4 \sqrt{6}} & \frac{1}{54} \\
 b_{\text{h3}}^2 b_{\text{v2}} & \frac{2 |001\rangle +\left(-1-i \sqrt{3}\right) |010\rangle }{2
   \sqrt{2}} & \frac{1}{54} \\
 b_{\text{h3}}^2 b_{\text{v3}} & \frac{i (|001\rangle +|010\rangle )}{\sqrt{2}} & \frac{1}{54}
   \\
 b_{\text{h3}} b_{\text{v1}}^2 & \frac{\left(3+i \sqrt{3}\right) |011\rangle -i \left(4 \sqrt{3}
   |110\rangle -\left(\sqrt{3}+3 i\right) |101\rangle \right)}{4 \sqrt{3}} & \frac{1}{27} \\
 b_{\text{h3}} b_{\text{v2}}^2 & \frac{1}{4} \left(-\left(\sqrt{3}-i\right) |011\rangle
   +\left(\sqrt{3}+i\right) |101\rangle -4 i |110\rangle \right) & \frac{1}{27} \\
 b_{\text{h3}} b_{\text{v3}}^2 & \frac{1}{2} i (|011\rangle +|101\rangle +2 |110\rangle ) &
   \frac{1}{27} 
\end{eqnarray*}

And finally, if we allow for three clicks in the same detector
\begin{eqnarray*}
 \text{\underline{Detection Pattern}} & \text{\underline{State}} & \text{\underline{Probability}} \\
 b_{\text{h1}}^3 & -|000\rangle  & \frac{1}{36} \\
 b_{\text{h2}}^3 & -|000\rangle  & \frac{1}{36} \\
 b_{\text{h3}}^3 & i |000\rangle  & \frac{1}{36} \\
 b_{\text{v1}}^3 & -|111\rangle  & \frac{1}{36} \\
 b_{\text{v2}}^3 & -|111\rangle  & \frac{1}{36} \\
 b_{\text{v3}}^3 & i |111\rangle  & \frac{1}{36} \\
\end{eqnarray*}
The probability of generating entanglement between three atoms when only non-number-resolved detectors are available is proportional to $(1/4)\eta^3$. When number-resolved detectors are available, this probability goes up to $(3/4)\eta^3$.

The following detection patterns are Hong-Ou-Mandel suppressed
\begin{eqnarray*}
 b_{\text{h1}}^2 b_{\text{h2}} \\
 b_{\text{h1}}^2 b_{\text{h3}} \\
 b_{\text{h1}} b_{\text{h2}}^2 \\
 b_{\text{h1}} b_{\text{h3}}^2 \\
 b_{\text{h2}}^2 b_{\text{h3}} \\
 b_{\text{h2}} b_{\text{h3}}^2 \\
 b_{\text{v1}}^2 b_{\text{v2}} \\
 b_{\text{v1}}^2 b_{\text{v3}} \\
 b_{\text{v1}} b_{\text{v2}}^2 \\
 b_{\text{v1}} b_{\text{v3}}^2 \\
 b_{\text{v2}}^2 b_{\text{v3}} \\
 b_{\text{v2}} b_{\text{v3}}^2 \\
\end{eqnarray*}

\section{\textit{n}-to-\textit{n} Symmetric Multiport}
\label{app:16Node}

\begin{figure}
\centerline{\includegraphics[width=0.5\columnwidth]{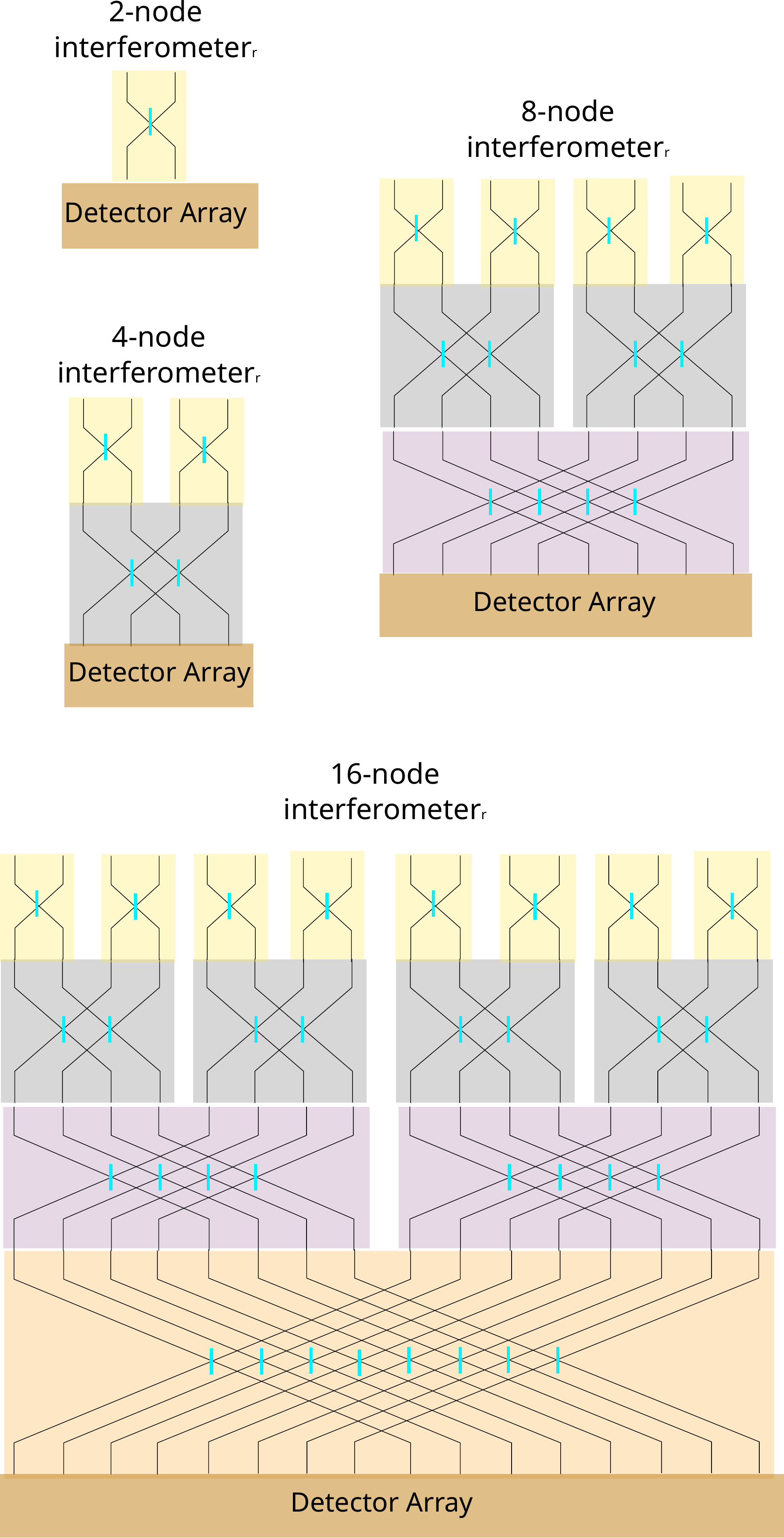}}
\caption{\(2^d\)-to-\(2^d\) multiport interferometer for \(n = 2, 4, 8,\) and \(16\).}
\label{fig:16inter}
\end{figure}

Figure \ref{fig:16inter} shows how to construct an \(n\)-to-\(n\) symmetric multiport for \(n=2^d\). Such a multiport can be used to entangle up to \(n=2^d\) nodes, including any subnetwork composed of \(m\) = 2, 3, 4, ..., \(n-1\) nodes, using either the entanglement-swapping scheme or the WPE scheme. The number of 50:50 beam splitters that each photon crosses is \(d\). Although the complexity of the interferometer increases with the number of input and output ports, when using the entanglement-swapping approach, a photon never interferes with itself, thereby maintaining the resilience of the scheme to path-length fluctuations. These kinds of interferometers can be built using bulk optics, although alignment and robustness are challenging. Fibre-based networks and photonic integrated circuits (PICs) are therefore more suitable.

\begin{figure}
\centerline{\includegraphics[width=0.5\columnwidth]{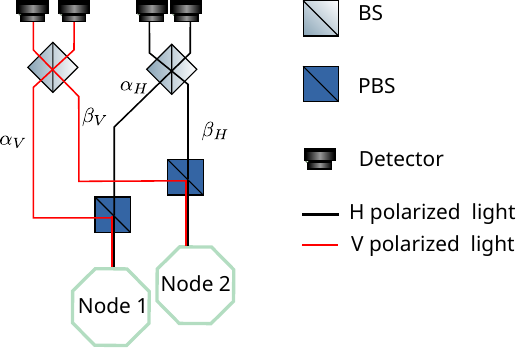}}
\caption{2-to-2 interferometer with separated polarisations. The H and V polarisation components of the photon are separated and sent to two independent interferometers.}
\label{fig:separate_interferometer}
\end{figure}

When using PICs and photon-polarisation encoding, a technical challenge encountered is the difficulty in building polarisation-preserving waveguides for two orthogonal polarisations, as well as polarisation-agnostic beam splitters. This can be overcome by separating the two incoming photon polarisations into two separate interferometers, one for H-polarisation and another for V-polarisation, as illustrated in Fig. \ref{fig:separate_interferometer} for the case of a 2-to-2 interferometer. This scheme still works because H and V photons never interfere anyway. Entanglement is heralded by the detection, in coincidence, of one H-photon and one V-photon in any of the outputs. The probability of entanglement generation is the same as using the standard setup (Fig. \ref{fig:GBSA_2_3_4}a), i.e., \(\propto \frac{1}{2}\eta^2\). The main difference is that now the phase of the created state is sensitive to the H and V path differences. Depending on which detectors click, we create either
\begin{align}
    \frac{1}{2}(\ket{01} \pm e^{i\phi}\ket{10}),
\end{align}
with $\phi = 2\pi\left[(\beta_H - \beta_V) - (\alpha_H - \alpha_V)\right]/\lambda$, where \(\alpha_{H,V}\) and \(\beta_{H,V}\) are the path lengths of the interferometers as illustrated in Fig. \ref{fig:separate_interferometer}, and $\lambda$ is the wavelength of the photons. Using temperature-stabilised PICs, achieving the required phase stability is not an issue, making this a viable option. The concept can be extended to any Bell state analyser, for any number of nodes, where the two polarisation modes can be separated into two different interferometers.

\end{document}